\documentclass[showpacs,preprintnumbers,
superscriptaddress,amsmath]{revtex4}

\newcommand{\NPB}[3]{Nucl.\ Phys.\ {\bf B#1},\ #2 (#3)}

\newcommand{\PLB}[3]{Phys.\ Lett.\ B\ {\bf #1},\ #2 (#3)}

\newcommand{\PRL}[3]{Phys.\ Rev.\ Lett.\ {\bf #1},\ #2 (#3)}

\newcommand{\PRD}[3]{Phys.\ Rev.\ D\ {\bf #1},\ #2 (#3)}

\renewcommand\a{\alpha}
\renewcommand\b{\beta}
\newcommand\g{\gamma}
\renewcommand\d{\delta}
\newcommand\e{\epsilon}
\newcommand\z{\zeta}

\renewcommand\k{\kappa}
\renewcommand\l{\lambda}
\newcommand\m{\mu}
\newcommand\n{\nu}
\newcommand\x{\xi}
\newcommand\p{\pi}

\newcommand\s{\sigma}
\renewcommand\t{\tau}

\newcommand{\be}{\begin{equation}}
\newcommand{\ee}{\end{equation}}
\newcommand{\bea}{\begin{eqnarray}}
\newcommand{\eea}{\end{eqnarray}}
\newcommand{\ba}[1]{\begin{array}{#1}}
\newcommand{\ea}{\end{array}}

\newcommand{\bm}[1]{\mbox{\boldmath${#1}$}}
 
\newcommand{\uk}{\hat{\mathbf{k}}}

\newcommand{\vg}{\bm{\gamma}}
\newcommand{\gperp}{\bm{\gamma}_{\perp}}

\begin{document}

\title{Mixing and Screening of Photons and Gluons in a Color 
Superconductor} 

\author{Andreas Schmitt}
\email{aschmitt@th.physik.uni-frankfurt.de}
\affiliation{Institut f\"ur Theoretische Physik, 
J.W. Goethe-Universit\"at, D-60054 Frankfurt/Main, Germany}

\author{Qun Wang}
\email{qwang@th.physik.uni-frankfurt.de}
\affiliation{Institut f\"ur Theoretische Physik, 
J.W. Goethe-Universit\"at, D-60054 Frankfurt/Main, Germany}
\affiliation{Physics Department, Shandong University, Jinan, 
Shandong, 250100, P.R. China}

\author{Dirk H. Rischke}
\email{drischke@th.physik.uni-frankfurt.de}
\affiliation{Institut f\"ur Theoretische Physik, 
J.W. Goethe-Universit\"at, D-60054 Frankfurt/Main, Germany}

\date{\today}

\begin{abstract}
We calculate the Debye and Meissner masses of photons and gluons
for a spin-one color superconductor in the polar and color-spin-locked
phases as well as for a spin-zero color
superconductor in the 2SC and color-flavor-locked phases. 
A general derivation for photon-gluon mixing is provided in terms 
of the QCD partition function. We show
qualitatively and quantitatively which of the gauge bosons attain a mass via 
the Anderson-Higgs mechanism.
Contrary to the spin-zero phases, both 
spin-one phases exhibit an electromagnetic Meissner effect. 
\end{abstract}

\maketitle

\section{Introduction} \label{intro}

Sufficiently cold and dense quark matter is a color 
superconductor. Due to asymptotic freedom, one-gluon exchange is the 
dominant interaction in a dense quark system. 
The one-gluon exchange interaction is attractive in the color-antitriplet
channel. Similar to electron Cooper pairing in ordinary superconductors 
\cite{bcs}, this leads to the formation of Cooper pairs at the quark Fermi 
surface \cite{bailin,alford1} (for a review, see \cite{review}). 
Cold and dense matter could occur
in the interior of compact stellar objects, such as neutron stars \cite{stars}.

In ordinary superconductivity, because of spontaneous
symmetry breaking of the electromagnetic gauge group $U(1)_{em}$, the
photon attains a mass via the Anderson-Higgs mechanism. Physically, 
this mass corresponds to screening of the magnetic field 
in superconducting matter. One of the consequences is the electromagnetic
Meissner effect, i.e., the expulsion of magnetic fields. In order to 
investigate the Anderson-Higgs mechanism in a color superconductor, one has 
to take into account the color gauge group $SU(3)_c$ in addition to the 
electromagnetic $U(1)_{em}$. The symmetry breaking pattern 
of $SU(3)_c\times U(1)_{em}$ determines which gauge bosons (gluons and 
photons) attain a mass. The order parameter for quark Cooper pairing 
defines the specific symmetry breaking pattern. In this paper we provide
a general treatment of the influence of color superconductivity on color 
and electromagnetic fields and apply it to four different phases of 
superconducting quark matter. 

First, we consider a system of two quark flavors. Here, two quarks can 
form Cooper pairs in the color-antitriplet, flavor-singlet channel 
\cite{bailin},
commonly called the 2SC phase. The Cooper pairs in this phase carry total spin 
zero, $J=0$. Moreover, in the 2SC phase the quarks of one color are 
supposed to remain unpaired.

Second, we study three quark flavors  
in the color-flavor-locked (CFL) phase \cite{alford2}. In this
phase, the condensate consists of pairs in the color-antitriplet, 
flavor-antitriplet channel. All colors and flavors participate in 
pairing and the condensate is 
invariant under joint color-flavor rotations. Also in this phase the Cooper
pairs carry spin zero.

In the third and fourth phases considered here, quarks form spin-one 
Cooper pairs, $J=1$ 
\cite{bailin,iwasaki,spin1}. For symmetry reasons, a total spin
$J=1$ is required for Cooper pairs of quarks which carry the same flavor.
The simplest case is a system of only one quark flavor. Another 
possibility is a many-flavor system where each flavor separately forms Cooper 
pairs. In the polar phase \cite{schaefer,schmitt}, the spin of the pair points 
in a fixed spatial direction and, as in the 2SC phase, one color remains 
unpaired. In the color-spin-locked (CSL) phase \cite{bailin}, each 
spatial direction is associated with a direction in color space. 
The case of spin-one Cooper pairing of quarks is in some respect similar to 
atomic pairing in superfluid Helium 3 \cite{vollhardt}. Therefore, 
similar phases occur in superconducting quark matter, e.g., the CSL phase 
corresponds to the ``B phase'' in Helium 3.        

Since the order parameter has a particular color-flavor-spin 
structure in each phase, the pattern of symmetry breaking differs from 
phase to phase, which consequently have to be investigated separately 
in order to identify the gauge bosons which 
attain a mass through the Anderson-Higgs mechanism. 
In some phases, namely the 2SC, CFL, and polar phases, it turns out that
the physical fields are 
not the original gluon and photon fields but linear
combinations of these, i.e., there is mixing between photons and gluons 
\cite{emprops,litim}.  The mixing angle is the analogue of the 
Weinberg angle in the standard model of electroweak interactions. 

Our paper is organized as follows. In Sec.\ \ref{phoglu} we discuss
the QCD partition function for massless quarks in the presence of 
gluon and photon gauge fields. We include a diquark condensation
term to account for color superconductivity. The goal of this 
section is to determine the gauge boson propagators. 
This fundamental study provides the 
basis for the following sections. It is general in the sense that 
we consider gauge fields for arbitrary four-momenta $P$ while 
afterwards we focus on the limiting case $P\to 0$ in order to discuss screening
of static and homogeneous electric and magnetic fields. 

In Sec.\ \ref{symmix}, we use a group-theoretical argument, namely
the invariance of the gap matrix under some subgroup of 
$SU(3)_c\times U(1)_{em}$, to show which of the gauge bosons attain 
a mass via the Anderson-Higgs mechanism. In this section we 
determine in a qualitative manner if electric and magnetic gauge 
fields are screened. 
Moreover, we derive the mixing angles describing
the rotation of the gauge fields. 

In the following two sections we extend the results of Sec.\ \ref{symmix}
by performing a quantitative calculation of the Debye and Meissner masses for
all gauge bosons. These masses correspond to the inverse screening lengths
for electric and magnetic fields, respectively. They are obtained 
from the longitudinal and transverse components of the polarization 
tensors in the zero-energy, low-momentum limit.
The first of these two sections, Sec.\ \ref{calc}, deals with the
technical details of the calculation. 
In its first part, Sec.\ \ref{generalpol}, we perform all calculations
that do not depend on the special color-superconducting phase. A part
of these calculations, 
namely the integrals over the absolute value of the quark momentum, 
are deferred to App.\ \ref{AppA}.  In the second part,
Secs.\ \ref{2SCphase} - \ref{CSLphase}, we specify the results for the above 
mentioned four color-superconducting phases, i.e., we derive the 
particular expressions for the relevant components of the polarization tensors.
Readers who are not interested in the technical details may skip this
section. All results and a discussion are given in Sec.\ \ref{results}.
We present all screening masses for photons and gluons which yield the
mixing angles and the masses for the (physically relevant)
mixed, or rotated, gauge bosons. One of the points we discuss in this 
section is in which cases these mixing angles differ from the ones 
obtained in Sec.\ \ref{symmix}. 
We conclude and summarize our study in Sec.\ \ref{summary}.  
       
Our convention for the metric tensor is 
$g^{\mu\nu}=\mbox{diag}\{1,-1,-1,-1\}$. 
Our units are $\hbar=c=k_B=1$. Four-vectors
are denoted by capital letters, 
$K\equiv K^\mu=(k_0,{\bf k})$, 
and $k\equiv|{\bf k}|$, while $\uk\equiv{\bf k}/k$.
We work in the imaginary-time formalism, i.e., $T/V \sum_K \equiv
T \sum_n \int d^3{\bf k}/(2\pi)^3$, where $n$ labels the Matsubara 
frequencies $\omega_n \equiv i k_0$. For bosons, $\omega_n=2n \pi T$,
for fermions, $\omega_n=(2n+1) \pi T$.

\section{Mixing of gluons and photons} \label{phoglu}

In this section we provide the theoretical basis for the calculation 
of the Debye and Meissner masses in a color superconductor. We clarify
how ``mixing'' of gluon and photon fields has to be understood. The results of
this section are the final form for gluon and photon propagators, 
Eq.\ (\ref{phogluprop}), and the mixing angle between gluon and photon
fields, Eq.\ (\ref{theta}).  

\subsection{Gluon and photon propagators} \label{propagators}

We start from the grand partition function of QCD for
massless quarks in the presence of gluon fields $A_a^\mu$ and
a photon field $A^\mu$,
\be
{\cal Z}=\int {\cal D}A\,e^{S_A}{\cal Z}_q[A] \,\, .
\ee 
Here, $S_A$ is the action for gluon and photon fields. It consists
of three parts,
\be
S_A=S_{F^2} + S_{gf} + S_{FPG} \,\, ,
\ee
where $S_{gf}$ and $S_{FPG}$ are the gauge fixing and 
ghost terms, respectively, and 
\be 
S_{F^2}\equiv-\frac{1}{4}\int_X (F_a^{\m\n}F^a_{\m\n}+F^{\m\n}F_{\m\n})
\ee
is the gauge field part. The space-time integration is defined
as $\int_X\equiv\int_0^{1/T}d\tau\int_V d^3{\bf x}$, where $T$
is the temperature and $V$ the volume of the system.
The field strength tensors $F^a_{\m\n}=
\partial_\m A_\n^a-\partial_\n A_\m^a + g f^{abc}A_\m^b A_\n^c$ 
correspond to the gluon fields $A_\m^a$, while $F_{\m\n}=
\partial_\m A_\n-\partial_\n A_\m$ corresponds to the photon 
field $A_\m$. The functional ${\cal Z}_q[A]$ is the grand
partition function of massless quarks in the presence of
gluon and photon fields and a chemical potential $\m$. It is given by
\be \label{quarkpartition}
{\cal Z}_q[A]=\int{\cal D}\bar{\psi}\,{\cal D}\psi\exp\left[\int_X
\bar{\psi}\,(i\,\gamma^\mu\partial_\mu+\mu\gamma_0+g\,\gamma^\mu A_\mu^aT_a
+e\,\gamma^\mu A_\mu Q)\, \psi\right] \,\, ,
\ee
where $T_a$ are the generators in the fundamental representation 
of the gauge group of strong
interactions, $SU(3)_c$, and $Q$ is the generator of the
electromagnetic $U(1)_{em}$. The coupling constants for the
strong interaction and electromagnetism are denoted by $g$ and
$e$, respectively. The quark fields $\psi$ are spinors in color, flavor,
and Dirac space. 

In order to take into account Cooper pairing of the quarks we 
include a diquark source term. This has been done for instance
in Refs.\ \cite{bailin,shovkovy}. We generalize this treatment by
also taking into account the photon field.
Then, after integrating out the fermion fields, the partition function
is given by \cite{shovkovy}
\be \label{log}
{\cal Z}=\int {\cal D}A\exp\left[S_A+\frac{1}{2}
{\rm Tr}\,\ln({\cal S}^{-1}+A_\mu^a\hat{\Gamma}^\mu_a)\right] \,\, ,
\ee
where the trace runs over space-time, Nambu-Gor'kov, color, flavor, and Dirac
indices.  The sum over $a$ now runs from 1 to 9, where $A_\mu^9\equiv A_\m$ 
is the photon field. We also defined the $2\times 2$ Nambu-Gor'kov matrices 
\be \label{vertex}
\hat{\Gamma}_a^\m\equiv {\rm diag}\left(\Gamma_a^\m,
\overline{\Gamma}_a^\m\right)
\equiv \left\{\begin{array}{lll} {\rm diag}(g\,\g^\m T_a,-g\,\g^\m T_a^T) & 
\mbox{for} & \quad a=1,\ldots,8 \,\, , \\ \\
{\rm diag}(e\, \g^\m Q,-e\,\g^\m Q) & \mbox{for} & \quad a=9 \,\, .
\end{array} \right.    
\ee
${\cal S}\equiv {\cal S}(X,Y)$ is the quasiparticle propagator in 
Nambu-Gor'kov space. In Eq.\ (\ref{log}), we did not explicitly keep the
fluctuations of the order parameter field, cf.\ Eq.\ (26) of Ref.\ 
\cite{shovkovy}. This is possible, since we are only interested in
the gluon and photon propagator. Nevertheless, it should be kept in mind
that, like in any theory with spontaneously broken gauge symmetry, 
these fluctuations mix with the (4-)longitudinal (unphysical) 
components of the gauge field. As usual, using a suitable choice of
't Hooft gauge, the fluctuations can be decoupled from the gauge field.
Then, in unitary gauge, one finds that the gauge boson propagator is explicitly
(4-)transverse \cite{shovkovy}, 
in accordance with general principles \cite{lee,lebellac}.

Expanding the logarithm in Eq.\ (\ref{log}) to second order in the 
gauge fields, we obtain
\be
\frac{1}{2}{\rm Tr}\,\ln({\cal S}^{-1}+A_\mu^a\hat{\Gamma}^\mu_a)
\simeq\frac{1}{2}{\rm Tr}\,\ln{\cal S}^{-1}
+\frac{1}{2}\,{\rm Tr}[A_\mu^a\,\hat{\Gamma}_a^\mu\,{\cal S}]
-\frac{1}{4}{\rm Tr}[A_\mu^a\,\hat{\Gamma}_a^\mu\,{\cal S}
\,A_\nu^b\,\hat{\Gamma}_b^\nu\,{\cal S}] \,\, .
\ee
Following Ref.\ \cite{shovkovy}, the sum of all terms
which are quadratic in the gauge fields will be denoted by $S_2$. 
$S_2$ does not only contain the 
pure gluon and photon terms but also two terms
which mix gluon and photon fields. In order to perform
the trace over space-time, we introduce the Fourier transforms
\begin{subequations}
\bea
{\cal S}(X,Y)&=&\frac{T}{V}\sum_K e^{-iK\cdot (X-Y)} {\cal S}(K) \,\, , \\
A_\mu^a(X)&=&\sum_P e^{-iP\cdot X} A_\m^a(P) \,\, ,
\eea
\end{subequations}
where we used translational invariance,
${\cal S}(X,Y)={\cal S}(X-Y)$. 
Then we obtain
\bea \label{S2}
S_2 & = & -\frac{1}{4}\int_{X,Y}{\rm Tr}[A_\mu^a(X)\,\hat{\Gamma}_a^\mu
\,{\cal S}(X,Y)\,A_\nu^b(Y)\,\hat{\Gamma}_b^\nu\,{\cal S}(Y,X)] \nonumber \\
& = & -\frac{1}{4}\sum_{K,P}{\rm Tr}[A_\mu^a(-P)\,\hat{\Gamma}_a^\mu
\,{\cal S}(K)\,A_\nu^b(P)\,\hat{\Gamma}_b^\nu\,{\cal S}(K-P)] \nonumber \\ 
& = & -\frac{1}{2}\frac{V}{T}\sum_P A_\mu^a(-P) \, \Pi_{ab}^{\m\n}(P)
\, A_\n^b(P) \,\, ,
\eea
where the trace now runs over Nambu-Gor'kov, color, flavor, and Dirac
indices and where we defined the polarization tensor
\be \label{poltensors} 
\Pi_{ab}^{\m\n}(P)\equiv\frac{1}{2}\frac{T}{V}\sum_K{\rm Tr}
[\hat{\Gamma}_a^\mu\,{\cal S}(K)\,\hat{\Gamma}_b^\nu\,{\cal S}(K-P)] \,\, .
\ee
The Nambu-Gor'kov quasiparticle propagator in momentum space is defined as
\be
{\cal S}(K)\equiv \left(\begin{array}{cc} G^+(K) & \Xi^-(K) \\ 
\Xi^+(K) & G^-(K) \end{array}\right) \,\, , 
\ee
with the quasiparticle propagators $G^\pm(K)$ and the ``anomalous''
propagators $\Xi^\pm(K)$, 
\be
G^\pm(K)=\left\{[G_0^\pm]^{-1}(K)-\Phi^\mp(K)\, G_0^\mp(K) \,\Phi^\pm(K)
\right\}^{-1} \quad,\qquad
\Xi^\pm(K) = -G_0^\mp(K)\Phi^\pm(K) G^\pm(K) \,\, .
\ee
Here, $G_0^\pm(K)=(\g_\m K^\m\pm\m\g_0)^{-1}$ is the free
propagator. In these expressions, we have put the regular self-energies
(cf.\ Ref.\ \cite{review}) 
to zero since, to the order we are computing, they do not influence 
the results for the polarization tensors.
Following Ref.\ \cite{schmitt}, 
we write the gap matrix $\Phi^+$ for color-superconducting quark matter, where 
ultrarelativistic quarks form Cooper pairs in the even-parity channel, as
\be \label{gapmatrix}	
\Phi^+(K)=\sum_{e=\pm}\phi^e(K){\cal M}_{\bf k}\Lambda_{\bf k}^e \,\, .
\ee
Here, $\phi^e(K)$ is the gap function, $\Lambda_{\bf k}^e=(1+e\g_0
\vg \cdot \uk)/2$ are projectors onto states of positive or 
negative energy, and ${\cal M}_{\bf k}$ is a matrix in color, flavor, and 
spin space that determines the color-superconducting phase. One can
choose ${\cal M}_{\bf k}$ such that $[{\cal M}_{\bf k},\Lambda_{\bf k}^e]=0$.
Furthermore, $\Phi^-(K)\equiv \g_0\left(\Phi^+(K)\right)^\dag\g_0$.
For the color-superconducting phases considered here, $\g_0 
{\cal M}_{\bf k}^\dagger {\cal M}_{\bf k} \g_0 = {\cal M}_{\bf k}
{\cal M}_{\bf k}^\dagger$. In this case, 
\be  \label{prop}
G^\pm(K)=[G_0^\mp(K)]^{-1}\sum_{e=\pm}\sum_{r=1,2}{\cal P}^r_{\bf k}
\, \Lambda_{\bf k}^{\mp e}\, \frac{1}{k_0^2-\left[\e_{k,r}^e(\phi^e)\right]^2}
\,\, ,
\ee
where ${\cal P}^r_{\bf k}$ are projectors onto the eigenspaces
of the hermitian matrix $\g_0{\cal M}_{\bf k}^\dag{\cal M}_{\bf k}\g_0$. In all
phases considered here, there are two eigenvalues of this matrix,
denoted by $\l_1$ and $\l_2$. They appear in the quasiparticle 
excitation energies
\be \label{excite}
\e_{k,r}^e(\phi^e)\equiv \left[(\m-ek)^2+\l_r|\phi^e|^2\right]^{1/2} \,\, .
\ee
The projectors are given by
\be \label{proj}
{\cal P}^{1,2}_{\bf k} = \frac{\g_0{\cal M}_{\bf k}^\dag{\cal M}_{\bf k}\g_0
-\l_{2,1}}{\l_{1,2}-\l_{2,1}} \,\, .
\ee
The anomalous propagators can be written as 
\be \label{aprop}
\Xi^+(K) = -\sum_{e,r}\g_0{\cal M}_{\bf k}\g_0 \, {\cal P}^r_{\bf k} \,
\Lambda_{\bf k}^{-e}\, \frac{\phi^e(K)}{k_0^2-\left[\e_{k,r}^e(\phi^e)
\right]^2}
\,\, , \qquad
\Xi^-(K) = -\sum_{e,r}{\cal M}_{\bf k}^\dagger \, {\cal P}^r_{\bf k} \,
\Lambda_{\bf k}^e\, \frac{\phi^{e\,*}(K)}{k_0^2-\left[\e_{k,r}^e(\phi^e)
\right]^2}\,\, .
\ee
As in Ref.\ \cite{shovkovy},  we introduce a set of complete, 
orthogonal projectors for each 4-vector $P^\m=(p_0,{\bf p})$,
\be \label{defABE}
{\cal Q}_1^{\m\n}\equiv g^{\m\n}-{\cal Q}_2^{\m\n}-{\cal Q}_3^{\m\n} 
\,\, , \qquad
{\cal Q}_2^{\m\n}\equiv\frac{N^\m N^\n}{N^2} \,\, , \qquad
{\cal Q}_3^{\m\n}\equiv\frac{P^\m P^\n}{P^2}\,\, .
\ee
With $N^\m\equiv(p_0p^2/P^2,p_0^2{\bf p}/P^2)$, the projector
${\cal Q}_2^{\m\n}$ projects onto the one-dimensional subspace that is 
(4-)orthogonal to $P^\m$ but (3-)parallel to ${\bf p}$. The
operator ${\cal Q}_3^{\m\n}$ projects onto the one-dimensional 
subspace parallel to $P^\m$.  
Consequently, ${\cal Q}_1^{\m\n}$ projects onto a two-dimensional subspace
that is (4-)orthogonal to both $P^\m$ and $N^\m$. Furthermore,
this subspace is (3-)orthogonal to ${\bf p}$. With the additional 
tensor
\be \label{defC}
{\cal Q}_4^{\m\n}=N^\m P^\n + P^\m N^\n
\ee
we can decompose the polarization tensor \cite{lebellac},
\be \label{decompose}
\Pi^{\m\n}_{ab}(P)=\sum_i\Pi^i_{ab}(P)\,{\cal Q}_i^{\m\n} \,\, .   
\ee
(In  the notation of Refs.\ \cite{shovkovy,lebellac}, ${\cal Q}_1\equiv 
{\rm A}, {\cal Q}_2\equiv {\rm B},{\cal Q}_3\equiv {\rm E},
{\cal Q}_4\equiv {\rm C} $.) 
Since for $i=1,2,3$, ${\cal Q}_i^{\m\n} {\cal Q}_{4\n\m}=0$, 
the coefficients $\Pi^i_{ab}(P)$ of the 
projection operators are given by 
\be \label{components}
\Pi^i_{ab}(P) = \frac{\Pi^{\m\n}_{ab}(P){\cal Q}_{i\n\m}}{{\cal Q}_{i\l}^\l} 
\quad, \qquad i=1,2,3 \,\, .
\ee
The remaining coefficient corresponding to ${\cal Q}_4$ is
\be \label{Pi4}
\Pi^4_{ab}(P) = \frac{ \Pi_{ab}^{\m\n}(P){\cal Q}_{4\n\m}}{{\cal Q}_4^{\l\s}
{\cal Q}_{4\s\l}}  \,\, .
\ee
The explicit forms for $\Pi^i_{ab}(P)$ are given
in Ref.\ \cite{shovkovy}, Eqs.\ (42) and (43). 
Employing the decomposition of the polarization tensor in Eq.\ (\ref{S2}),
we obtain
\bea 
S_2&=&-\frac{1}{2}\frac{V}{T}\sum_P\sum_i A_\mu^a(-P)
{\cal Q}_i^{\m\n}\Pi_{ab}^i(P)A_\n^b(P) \,\, .
\eea 
Now we add the free gauge field term
\be
S^{(0)}_{F^2}=-\frac{1}{2}\frac{V}{T}\sum_PA_\m^a(-P)
(P^2g^{\m\n}-P^\m P^\n)A_\n^a(P)=-\frac{1}{2}\frac{V}{T}\sum_P 
A_\m^a(-P)P^2({\cal Q}_1^{\m\n}+{\cal Q}_2^{\m\n})A_\n^a(P) \,\,.
\ee
We obtain
\be \label{before}
S_2 + S^{(0)}_{F^2} =
-\frac{1}{2}\frac{V}{T}\sum_P A_\m^a(-P)\left\{ \sum_{i=1}^2 {\cal Q}_i^{\m\n}
\left[\d_{ab}P^2 + \Pi_{ab}^i(P)\right] + \sum_{i=3}^4 {\cal Q}_i^{\m\n}
\Pi_{ab}^i(P)\right\} A_\n^b(P) \,\, .
\ee
Since we finally want to read off the gluon and photon propagators, we have
to transform this expression in two ways.
First, concerning the Dirac structure it is necessary to get rid of 
the term proportional to ${\cal Q}_4$ which mixes the longitudinal
mode (3-parallel to ${\bf p}$) with the unphysical mode (4-parallel to
$P^\m$). Then, inverting the inverse propagator becomes trivial, because
it is just a linear combination
of the complete, orthogonal projectors ${\cal Q}_1, {\cal Q}_2, {\cal Q}_3$.
Second, in order to obtain the physical modes we have to diagonalize 
the resulting $9\times 9$ matrices
which, after eliminating ${\cal Q}_4$, will replace  
$\d_{ab}P^2 + \Pi_{ab}^i(P)$, $i=1,2$, and 
$\Pi_{ab}^i(P)$, $i=3,4$ in Eq.\ (\ref{before}). 

We first write $S_2 + S^{(0)}_{F^2}$ as 
\bea \label{withoutgauge}
S_2 + S^{(0)}_{F^2} 
&=&-\frac{1}{2}\frac{V}{T}\sum_{P}\Big\{\sum_{i=1}^2 A_i^{a\m}(-P) 
\left[\d_{ab}P^2 + \Pi_{ab}^i(P)\right] A^b_{i\m}(P) + A_3^{a\m}(-P) 
 \Pi_{ab}^3(P) A^b_{3\m}(P) \nonumber \\
&& \hspace{2.1cm} + A_2^{a\m}(-P) N_\m \Pi^4_{ab}(P) P_\n A_3^{b\n}(P)
+A_3^{a\m}(-P) P_\m \Pi^4_{ab}(P) N_\n A_2^{b\n}(P) \Big\} \,\, ,
\eea
where $A_i^{a\m}(P)\equiv {\cal Q}_i^{\m\n} A_\n^a(P)$ are the gauge fields
projected on the subspace corresponding to ${\cal Q}_i$. 
Now one can ``unmix'' the  
fields $A_2^\m(P)$ and $A_3^\m(P)$ by the following transformation
of the unphysical field component $A^a_{3\m}(P)$,
which does not affect the final result since we integrate over all fields
in the partition function,
\begin{subequations} \label{shift}
\bea 
A_3^{a\m}(-P) &\longrightarrow& A_3^{a\m}(-P) - 
A_2^{c\n}(-P)N_\n \Pi^4_{cb}(P)\left[\Pi^3(P)\right]^{-1}_{ba}
\, P^\m \,\, ,\\
A^a_{3\m}(P) &\longrightarrow& A^a_{3\m}(P) - 
P_\m \left[\Pi^3(P)\right]^{-1}_{ab}
\Pi^4_{bc}(P)\,N_\n A_2^{c\n}(P) \,\, .
\eea
\end{subequations}
After this transformation,
one is left with quadratic expressions in the projected fields.
The transformation modifies the term corresponding to $i=2$, 
\bea 
S_2 + S^{(0)}_{F^2} &=& 
-\frac{1}{2}\frac{V}{T}\sum_P A_\m^a(-P)\Big({\cal Q}_1^{\m\n}
\left[\d_{ab}P^2 + \Pi_{ab}^1(P)\right] \nonumber \\ 
&& + {\cal Q}_2^{\m\n}
\left\{\d_{ab}P^2 + \Pi_{ab}^2(P)-P^2N^2\Pi^4_{ac}(P)\,
\left[\Pi^3(P)\right]^{-1}_{cd}\,\Pi^4_{db}(P)\right\} 
+{\cal Q}_3^{\m\n}
\Pi^3_{ab}(P)\Big) A^b_\n(P) \,\, . \label{S_2+S_F}
\eea
Before we do the diagonalization in the 9-dimensional 
gluon-photon space, we add the gauge fixing term $S_{gf}$. We choose the
following gauge with gauge parameter $\l$,
\be \label{gf}
S_{gf}=-\frac{1}{2\l}\frac{V}{T}\sum_P A_\m^a(-P)P^\m  P^\n A_\n^a(P)
=-\frac{1}{2\l}\frac{V}{T}\sum_P A_\m^a(-P)P^2{\cal Q}_3^{\m\n}
A_\n^a(P) \,\, .
\ee
This looks like a covariant gauge but, including the fluctuations 
of the order parameter, which we did not write explicitly, 
it is actually some kind of 't Hooft gauge, cf.\ Eq.\ (50) of
Ref.\ \cite{shovkovy}. Moreover, had we fixed the gauge {\em prior\/}
to the shift (\ref{shift}) of the gauge fields, we would have
to start with an expression which is non-local and also involves the 
(3-)longitudinal components $A_2^{a\m}$ of the gauge field.

Adding the gauge fixing term (\ref{gf}) to Eq.\ (\ref{S_2+S_F}) leads to 
\be
S_2 + S^{(0)}_{F^2} +S_{gf} =  
-\frac{1}{2}\frac{V}{T}\sum_P \sum_{i=1}^3 A^a_\m(-P){\cal Q}_i^{\m\n}
\Theta^i_{ab}(P)A^b_\n(P) \,\, ,
\ee
with
\be \label{thetahat}
\Theta^i_{ab}(P) \equiv \left\{ \begin{array}{cl}
\d_{ab}P^2 + \Pi_{ab}^1(P)  & 
\mbox{for}\quad i=1  \,\, ,
\\ \\
\d_{ab}P^2 + \Pi_{ab}^2(P)-P^2N^2\Pi^4_{ac}(P)\,
\left[\Pi^3(P)\right]^{-1}_{cd}\,\Pi^4_{db}(P) & \mbox{for}\quad i=2 \,\, ,
\\ \\
\d_{ab}\frac{1}{\l}P^2+\Pi^3_{ab}(P) & \mbox{for}\quad i=3 \,\, .
\end{array} \right. 
\ee

In order to obtain the physical modes, we have to diagonalize 
the $9\times 9$ matrices $\Theta^i_{ab}(P)$. Since 
$\Theta^i_{ab}(P)$ is real and symmetric, 
diagonalization is achieved via an orthogonal 
transformation with a $9\times 9$ matrix ${\cal O}_i(P)$, 
\be \label{rotatedfields}
S_2 + S^{(0)}_{F^2} + S_{gf}=-\frac{1}{2}\frac{V}{T}\sum_P\sum_{i=1}^3      
\tilde{A}_{\mu,i}^a(-P){\cal Q}_i^{\m\n}
\tilde{\Theta}_{aa}^i(P)\tilde{A}_{\n,i}^a(P) \,\, ,
\ee
where 
\be
\tilde{A}_{\mu,i}^a(P)={\cal O}_i^{ab}(P)A_\m^b(P)
\ee
are the rotated gauge fields and
\be
\tilde{\Theta}_{aa}^i(P)={\cal O}_i^{ab}(P)
\Theta^i_{bc}(P){\cal O}_i^{ac}(P) 
\ee
are diagonal matrices.
The index $i$ in $\tilde{A}^a_{\m,i}$
has a different origin than in $A^a_{i\m}$ introduced in 
Eq.\ (\ref{withoutgauge}). For $\tilde{A}^a_{\m,i}$, it indicates 
that for each $i=1,2,3$ one has to perform a separate diagonalization. 
For $A^a_{i\m}$ it denoted the
projection corresponding to the projector ${\cal Q}_i^{\m\n}$. 

Note that the orthogonal matrix ${\cal O}_i(P)$ depends on $P^\m$.
For energies and momenta much larger than the superconducting gap parameter
the polarization tensor is explicitly (4-)transverse, $\Pi^3=0$,
and diagonal in $a, b$. In this case, ${\cal O}_i(P)\to {\bf 1}$. Consequently,
the gauge fields are not rotated. However, in the limit $p_0=0$, $p\to 0$
it is known that gluons and photons mix at least in the 2SC and CFL
phases \cite{emprops}, ${\cal O}_i(P)\neq {\bf 1}$. 
Thus, the mixing angle between
gluons and photons, which will be discussed
in Sec.\ \ref{mixing}, is in general a function of $P^\m$ and interpolates
between a nonvanishing value at $p_0=0$, $p\to 0$ and zero when $p_0,p\to
\infty$. Note also that the orthogonal matrix ${\cal O}_i(P)$ depends on
$i$, i.e., it may be different for longitudinal and transverse modes. We
comment on this in more detail below.

{}From Eq.\ (\ref{rotatedfields}), we can immediately read off the inverse 
propagator for gluons and photons. It is
\be
{\Delta^{-1}}_{aa}^{\m\n}(P) =
\sum_{i=1}^3{\cal Q}_i^{\m\n}\tilde{\Theta}_{aa}^i(P) \,\, .
\ee
{}From the definition of $\tilde{\Theta}_{aa}^i(P)$ we conclude
\be  \label{diaglimit}
\tilde{\Theta}^i_{aa}(P) \equiv \left\{ \begin{array}{cl} P^2+
\tilde{\Pi}_{aa}^i(P) & \mbox{for}\quad i=1,2  \,\, ,
\\ \\
\frac{1}{\l}P^2+\tilde{\Pi}^3_{aa}(P) & \mbox{for}\quad i=3 \,\, ,
\end{array} \right. 
\ee
where 
\be \label{diag}
\tilde{\Pi}_{aa}^i(P)=\left\{ \begin{array}{cl}{\cal O}_i^{ab}(P)
\Pi^i_{bc}(P){\cal O}_i^{ac}(P) & \mbox{for}\quad i=1,3  \,\, ,
\\ \\
{\cal O}_2^{ab}(P)\left\{
\Pi_{bc}^2(P)-P^2N^2\Pi^4_{bd}(P)\,
\left[\Pi^3(P)\right]^{-1}_{de}\,\Pi^4_{ec}(P)\right\}{\cal O}_2^{ac}(P) 
& \mbox{for}\quad i=2 \,\, .
\end{array} \right.  
\ee
In the case $p_0=0$, using Eqs.\ (42) and (43)
of Ref.\ \cite{shovkovy} one realizes that
the extra term involving $\Pi^3$ and $\Pi^4$ 
for $i=2$ vanishes. Thus, in this case
one only has to diagonalize the original polarization tensors $\Pi^i_{ab}$.

Finally, we end up with the gauge boson propagator 
\be \label{phogluprop}
\Delta_{aa}^{\m\n}(P)=\frac{1}{P^2+\tilde{\Pi}_{aa}^1(P)}{\cal Q}_1^{\m\n}+ 
\frac{1}{P^2+\tilde{\Pi}_{aa}^2(P)}{\cal Q}_2^{\m\n}+ 
\frac{\l}{P^2+\l\tilde{\Pi}_{aa}^3(P)}{\cal Q}_3^{\m\n} \,\, .
\ee
Setting the gauge parameter $\l=0$, we are left with the transverse and 
(3-)longitudinal modes, in accordance with general principles
\cite{lee,lebellac}. The static color and electromagnetic properties
of the color superconductor are characterized by the Debye masses
$\tilde{m}_{D,a}$ and the Meissner masses $\tilde{m}_{M,a}$ which are 
defined as
\begin{subequations} \label{defmasses}
\bea 
\tilde{m}^2_{D,a}&\equiv& -\lim\limits_{p\to 0}\tilde{\Pi}_{aa}^2(0,{\bf p})
=-\lim\limits_{p\to 0}\tilde{\Pi}_{aa}^{00}(0,{\bf p}) \,\, ,\\
\tilde{m}^2_{M,a}&\equiv& -\lim\limits_{p\to 0}\tilde{\Pi}_{aa}^1(0,{\bf p})
=\frac{1}{2}\lim\limits_{p\to 0}
(\d^{ij}-\hat{p}^i\hat{p}^j)\tilde{\Pi}_{aa}^{ij}(0,{\bf p})
 \,\, . 
\eea
\end{subequations} 
Since the orthogonal matrices ${\cal O}_i(P)$ are regular in the limit 
$p_0= 0$, $p\to 0$, the masses can also be obtained by first computing
$\lim_{p\to 0}\Pi_{ab}^{\m\n}(0,{\bf p})$ and then
diagonalizing the resulting $9\times 9$ mass matrix. In Sec.\
\ref{calc} we use this method to compute $\tilde{m}^2_{D,a}$
and $\tilde{m}^2_{M,a}$, since the diagonalization of the matrix
$\Pi_{ab}^{\m\n}(P)$ for arbitrary $P^\m$ is too difficult.

\subsection{The mixing angle} \label{mixing}

In this section we investigate the structure of the orthogonal 
matrices ${\cal O}_i(P)$ which diagonalize the gauge field part of
the grand partition function. 
In general, the matrices ${\cal O}_i(P)$ mix all gluon components among 
themselves and with the photon. However, in the limit $p_0=0$, $p\to 0$
it turns out that 
the only non-zero off-diagonal elements of the tensor
$\Pi_{ab}^i\equiv \lim_{p\to 0} \Pi_{ab}^i(0,{\bf p})$ are 
$\Pi_{89}^i=\Pi_{8\g}^i=\Pi_{\g 8}^i$. 
Physically speaking, gluons do not mix among themselves and 
only the eighth gluon mixes with the photon. In this case, Eq.\ (\ref{diag})
reduces to the diagonalization of a $2\times 2$ matrix.   
Consequently, the diagonalization is determined by only one
parameter $\theta_i$ and the (nontrivial part of the) transformation 
operator reads
\be
{\cal O}_i=\left(\begin{array}{cc}
\cos\theta_i&\sin\theta_i\\
-\sin\theta_i&\cos\theta_i\end{array}\right) \,\, .
\ee
The new fields are
\begin{subequations} \label{newfields}
\bea 
\tilde{A}^8_{\m,i}&=&\cos\theta_i\,A^8_\m
+\sin\theta_i\,A_\m \,\, ,\\
\tilde{A}_{\m,i}&=&-\sin\theta_i\,A^8_\m
+\cos\theta_i\,A_\m 
\,\, .
\eea
\end{subequations}
The new polarization functions (eigenvalues of $\Pi_{ab}^i$) are 
\begin{subequations} \label{eigenval}
\bea
\tilde{\Pi}_{88}^i&=&\Pi_{88}^i\cos^2\theta_i+2\Pi_{8\g}^i
\sin\theta_i\cos\theta_i+\Pi^i_{\g\g}\sin^2\theta_i \,\, ,\\
\tilde{\Pi}^i_{\g\g}&=&\Pi_{88}^i\sin^2\theta_i-2\Pi_{8\g}^i
\sin\theta_i\cos\theta_i+\Pi^i_{\g\g}\cos^2\theta_i \,\, .
\eea
\end{subequations}
The mixing angle $\theta_i$ is given by
\be \label{theta}
\tan 2\theta_i=\frac{2\Pi_{8\g}^i}{\Pi_{88}^i-\Pi^i_{\g\g}} \,\, .
\ee
If $[\Pi_{8\g}^i]^2=\Pi_{88}^i\Pi^i_{\g\g}$, the determinant
of $\Pi_{ab}^i$ is zero, which means that there is a vanishing eigenvalue.
 In this case, we have
\be \label{cossintheta}
\cos^2\theta_i=\frac{\Pi_{88}^i}{\Pi_{88}^i+\Pi_{\g\g}^i} \,\, ,
\ee
and the new polarization tensors, Eqs.\ (\ref{eigenval}),
have the simple form 
\be \label{detvanish}
\tilde{\Pi}_{88}^i=\Pi_{88}^i+\Pi^i_{\g\g} \quad ,\qquad 
\tilde{\Pi}^i_{\g\g}=0 \,\, .
\ee
Physically, the vanishing polarization tensor for the rotated 
photon corresponds to the absence of the Meissner effect for $i=1$, 
or the absence of Debye screening for $i=2$.

\section{Symmetries and mixing} \label{symmix}

In Section \ref{mixing} we have discussed the mixing
angle $\theta_i$ that combines the electromagnetic field and the eighth 
gluon. The index $i$ refers to 
the projectors ${\cal Q}_i^{\m\n}$ and thus, in general, the mixing 
angle could be different for transverse modes, $i=1$, and longitudinal modes,
$i=2$. 

In this section, we present a method 
to determine the mixing angles via a simple group-theoretical 
consideration. 
As we know from ordinary superconductors, the non-zero 
electric charge of a Cooper pair leads to the Meissner effect and
thus corresponds to a non-vanishing Meissner mass $m_{M,\g}$ for the 
photon. Besides this magnetic screening there is also  
electric screening of photons described by the photon Debye mass 
$m_{D,\g}$. Of course, in a color superconductor, in addition to the 
electric charge we also have to take into account the color charge.
The group-theoretical method applied in the following
allows us to investigate whether there exists a (new) charge which generates
an unbroken symmetry. In this case, a Cooper pair is neutral with respect 
to this charge, and consequently one expects that there is
neither a Meissner effect nor Debye screening.
The new charge is a linear combination of electric and color charges.
Correspondingly, the associated gauge field is a linear combination of the 
photon and the eighth gluon field, which, in turn, defines the mixing
angle. The group-theoretical method only allows to identify a new charge, 
and thus does not distinguish between electric Debye or magnetic Meissner 
screening. Consequently, the mixing angles for longitudinal (electric) and 
transverse (magnetic) modes deduced by this method are identical, 
$\theta_1=\theta_2\equiv\theta$.

A Cooper pair is neutral with respect to a charge, if the gap matrix 
$\Phi$ is invariant
under a transformation generated by an operator $\tilde{Q}$ corresponding 
to this charge. In general, the gap matrix is a matrix in color, flavor, 
and spin space,
\be
\Phi=\phi_{ijk}\,e_i^c\otimes e_j^f \otimes e_k^J \,\, ,
\ee
where $\phi_{ijk}$ is the order parameter and 
$e_i^c\otimes e_j^f\otimes e_k^J$ is a basis for the 
color$\otimes$flavor$\otimes$spin representation of the group
$SU(3)_c\times SU(N_f)_f\times SU(2)_J$.
Since $\tilde{Q}$ generates a subgroup of 
the product group of strong and electromagnetic interactions,
$SU(3)_c\times U(1)_{em}$, it can be found by the following
invariance condition,
\be
\Phi\stackrel{!}{=}(g_c\times g_{em})\Phi(g_c^T\times g_{em}^T) \,\, ,
\ee
where $g_c\in SU(3)_c$ and $g_{em}\in U(1)_{em}$ act on $e_i^c$ and
$e_j^f$, respectively. With $g_c=\exp(i\k_aT_a)$, $g_{em}=\exp(i\k Q)$,
for infinitesimal transformations this condition can be written as
\be \label{subgroup}
0=\phi_{ijk}\left[i\k_a\left(T_a e_i^c+e_i^c T_a^T\right)\otimes
e_j^f\otimes e_k^J + e_i^c\otimes i\k\left(Q\,e_j^f+e_j^fQ^T\right)
\otimes e_k^J\right] \,\, ,
\ee           
and finding the charge $\tilde{Q}$ is equivalent to determining
the real coefficients $\k_a$, $\k$, where $a=1,\ldots,8$. 
Using the representations and the order parameters for
each phase as given in Table \ref{tablesymmetry} we obtain 
\begin{subequations} \label{subgroup2}
\bea 
\mbox{2SC:}\qquad 0&=&i\,\k_a\left(T_a e_3^c+e_3^c T_a^T\right)\otimes
e^f\otimes e^J + i\,\k\,e_3^c\otimes \left(Q\,e^f+e^fQ\right)
\otimes e^J \,\, , \\
\mbox{CFL:}\qquad 0&=&i\,\k_a\left(T_a e_j^c+e_j^c T_a^T\right)\otimes
e_j^f\otimes e^J + i\,\k\,e_j^c\otimes \left(Q\,e_j^f+e_j^fQ\right)
\otimes e^J \,\, , \\
\mbox{CSL:}\qquad 0&=&i\,\k_a\left(T_a e_j^c+e_j^c T_a^T\right)\otimes
e^f\otimes e_j^J + 2\,i\,\k\,q\,e_j^c\otimes e^f
\otimes e_j^J \,\, , \\
\mbox{polar:}\qquad 0&=&i\,\k_a\left(T_a e_3^c+e_3^c T_a^T\right)\otimes
e^f\otimes e_3^J + 2\,i\,\k\,q\,e_3^c\otimes e^f
\otimes e_3^J \,\, .
\eea           
\end{subequations}
With the special form of the charge generator $Q$, given in the 
third column of 
Table \ref{tablesymmetry}, Eqs.\ (\ref{subgroup2}) yield conditions
for the coefficients $\k_a$ and $\k$,
\begin{subequations} 
\bea 
\mbox{2SC:}&\quad& \k_1,\k_2,\k_3 \,\, \mbox{arbitrary}, \,\,\,
 \k_4=\ldots=\k_7=0, \,\, \k_8=-\frac{1}{\sqrt{3}}\k \,\, , \\
\mbox{CFL:}&\quad& \k_1=\ldots=\k_7=0, \,\, \k_8=\frac{2}{\sqrt{3}}\k 
\,\, , \\
\mbox{CSL:}&\quad& \k_1=\ldots=\k_8=\k=0 \,\, , \\
\mbox{polar:}&\quad& \k_1,\k_2,\k_3 \,\, \mbox{arbitrary}, \,\,\,
 \k_4=\ldots=\k_7=0, \,\, \k_8=-2\sqrt{3}\,q\,\k \,\, . 
\eea           
\end{subequations}
\begin{table}  
\begin{tabular}{|c||c|c|c|c|c|}
\hline
      & representation & order  & Q & generators of & $\eta$ 
\\ 
 & & parameter & & residual group & 
\\ \hline\hline
2SC   & $\,\,{\bf\bar{3}}_c\otimes{\bf 1}_f\otimes{\bf 1}_J\,\,$ & $\d_{i3}$ &
      diag$(2/3,-1/3)$ & $T_1,T_2,T_3,Q+\eta T_8$ & $-1/\sqrt{3}$ 
\\ \hline
CFL   & ${\bf\bar{3}}_c\otimes{\bf\bar{3}}_f\otimes{\bf 1}_J$ & $\d_{ij}$ &
      \,\, diag$(-1/3,-1/3,2/3) \,\, $ & $Q+\eta T_8$ & $2/\sqrt{3}$ 
\\ \hline
CSL   & ${\bf\bar{3}}_c\otimes{\bf 3}_J$ & $\d_{ik}$ &
      $q$ & -- & -- 
\\ \hline 
polar & ${\bf\bar{3}}_c\otimes{\bf 3}_J$ & $\d_{i3}\d_{k3}$ &
      $q$ & $\,\, T_1,T_2,T_3,Q+\eta T_8 \,\, $ & $\,\, -2\sqrt{3}\, q \,\,$ 
\\ \hline
\end{tabular}
\caption{Representations, order parameters $\phi_{ijk}$, charge
generators $Q$, generators of the residual symmetry group, and 
coefficients $\eta$ for all color-superconducting phases considered here. 
For the CSL and polar phases, the electric charge of the quark is denoted
by $q$. The last two columns result from Eqs.\ (\ref{subgroup2}).}
\label{tablesymmetry}
\end{table}
The fact that $\k_1,\k_2,\k_3$ are arbitrary in the 2SC and polar phases
means that there is an unbroken $SU(2)_c$ symmetry under which the order 
parameter is invariant. This is obvious because in these phases Cooper 
pairs carry anti-blue color charge which is not seen by the generators 
of this $SU(2)_c$, since they
operate exclusively on the red-green subspace. The relation between 
$\k_8$ and $\k$ determines a new $\tilde{U}(1)$ symmetry, generated by
 \be \label{Qtilde}
\tilde{Q}=Q+\eta \, T_8 \,\, ,
\ee
where $\eta$ is given in the fifth column of Table \ref{tablesymmetry} for
the different cases considered here. In the CSL phase, the fact that all 
$\k$'s vanish means that there is no unbroken residual symmetry.

In order to deduce the mixing angle $\theta$ we rewrite the covariant 
derivative, $D_\m = \partial_\m - igA_\m^a T_a - ieA_\m Q$, in terms
of the new charge generator $\tilde{Q}$ and the  
linearly independent generator  
\be
\tilde{T}_8 = T_8 + \l Q \,\, ,
\ee
which belongs to the broken part of the group; $\l$ is a real 
constant to be determined below. We also have to replace
the gauge fields $A_\m^8$ and $A_\m$ by the rotated fields 
(\ref{newfields}) and the associated coupling constants $g$, $e$ by 
new coupling constants $\tilde{g}$, $\tilde{e}$. Thus, we demand the 
identity \cite{emprops}
\be
gA_\m^8 T_8 + eA_\m Q  = \tilde{g}\tilde{A}_\m^8\tilde{T}_8 
+ \tilde{e}\tilde{A}_\m \tilde{Q} \,\, .
\ee
Inserting Eqs.\ (\ref{newfields}) for the rotated fields and the definitions
of $\tilde{Q}$ and $\tilde{T}_8$ we determine   
\be
\tilde{g}=g\cos\theta \quad , \qquad \tilde{e} = e\cos\theta \quad , \qquad
\l=-\eta\frac{e^2}{g^2} \,\, ,
\ee
and the mixing angle
\be \label{allangles}
\cos^2\theta=\frac{g^2}{g^2+\eta^2 e^2} \,\, .
\ee
Since $g\gg e$, the mixing angle in all three cases is small. Thus,
according to Eqs.\ (\ref{newfields}), the gluon almost remains a gluon
and the photon almost remains a photon and therefore it is justified to
call the rotated gauge bosons the new gluon and the new photon.

In the 2SC, CFL, and polar phases, the 
new charge is neither Debye- nor Meissner-screened. From
the argument presented at the end of Section \ref{mixing} we therefore
expect that the polarization tensor for the rotated photon vanishes 
in the zero-energy, zero-momentum limit,
\be \label{zero}
\tilde{\Pi}_{\g\g}^{1,2}=0 \,\, .
\ee
As we shall see in the following section, this argument is not quite correct 
for the 2SC and polar phases, as it neglects the effect of the
unpaired blue quarks on the screening of electric and magnetic fields.
In the CSL phase, on the other hand, there is no residual 
$\tilde{U}(1)$ symmetry and we conclude that all gauge fields
experience the Meissner effect and Debye screening, since in this
case there does not exist any charge with respect to which a Cooper 
pair is neutral.
Concerning color screening, one concludes from Table \ref{tablesymmetry} that
in the 2SC phase and the polar phase only the gluons 4-8 are
screened while in the CFL and CSL phases all gluons are screened.

\section{Calculation of the polarization tensors} \label{calc}

In this section, we calculate the polarization tensors 
$\Pi_{ab}^{\m\n}(P)$ given in Eq.\ (\ref{poltensors}) in the limit 
$p_0=0,p\to 0$. In this case, the Debye and Meissner masses, 
cf.\ Eqs.\ (\ref{defmasses}), are obtained from the coefficients 
of the first two projectors in the decomposition (\ref{decompose}). 
They will be calculated in the next section.
Here, we first derive a general expression for the polarization tensor 
that holds for all different phases. Then we insert the order parameters
of the 2SC, CFL, polar, and CSL phases, and show the results for
each phase separately.

\subsection{General structure of $\Pi_{ab}^{\m\n}(0)$} \label{generalpol}

We start from Eq.\ (\ref{poltensors}) and first perform the trace
over Nambu-Gor'kov space,
\bea \label{nambu}
\Pi_{ab}^{\m\n}(P)&=&\frac{1}{2}\frac{T}{V}\sum_K
\left\{{\rm Tr}[\Gamma_a^\m \, G^+(K) \, \Gamma_b^\n G^+(K-P)]
+{\rm Tr}[\overline{\Gamma}_a^\m \, G^-(K) \, \overline{\Gamma}_b^\n \, G^-(K-P)]
\right. 
\nonumber \\
&&\left.+{\rm Tr}[\Gamma_a^\m \, \Xi^-(K) \, \overline{\Gamma}_b^\n \, 
\Xi^+(K-P)]
+{\rm Tr}[\overline{\Gamma}_a^\m \, \Xi^+(K)\, \Gamma_b^\n \, \Xi^-(K-P)] 
\right\} \,\, ,
\eea
where the traces now run over color, flavor, and Dirac space. 
In the following we first consider the traces with the quark propagators
$G^\pm$ and afterwards investigate the traces containing the anomalous 
propagators $\Xi^\pm$. 

In order to find the results for the former, we first perform the
Matsubara sum. This is completely analogous to the calculation of
Ref.\ \cite{meissner2}. The only difference is our more compact
notation with the help of the projectors ${\cal P}_{\bf k}^r$, cf.\
Eq.\ (\ref{prop}). Thus, abbreviating $K_1\equiv K$, $K_2\equiv K-P$, and
$k_i\equiv |{\bf k}_i|$ for $i=1,2$, we conclude 
\begin{subequations} \label{matsub1}
\bea 
T\,\sum_{k_0}{\rm Tr}
\left[\Gamma_a^\m G^+(K_1)\Gamma_b^\n G^+(K_2)\right] 
&=&\sum_{e_1,e_2}\sum_{r,s}{\rm Tr}\left[\Gamma_a^\m\,\g_0\,
{\cal P}_{{\bf k}_1}^r\,\Lambda_{{\bf k}_1}^{-e_1} \,  
\Gamma_b^\n \,\g_0\,{\cal P}_{{\bf k}_2}^s\,
\Lambda_{{\bf k}_2}^{-e_2}\right]\, v_{e_1e_2}^{rs,+}
(k_1,k_2,p_0) \,\, , \\
T\,\sum_{k_0}{\rm Tr}
\left[\overline{\Gamma}_a^\m G^-(K_1)\overline{\Gamma}_b^\n G^-(K_2)\right] 
&=&\sum_{e_1,e_2}\sum_{r,s}{\rm Tr}\left[\overline{\Gamma}_a^\m\,\g_0\,
{\cal P}_{{\bf k}_1}^r\,\Lambda_{{\bf k}_1}^{e_1} \,  
\overline{\Gamma}_b^\n \,\g_0\,{\cal P}_{{\bf k}_2}^s\,
\Lambda_{{\bf k}_2}^{e_2}\right]\, v_{e_1e_2}^{rs,-}
(k_1,k_2,p_0) \,\, ,
\eea
\end{subequations}
where (cf.\ Eq.\ (40) of Ref.\ \cite{meissner2})
\begin{subequations} \label{defv}
\bea 
v_{e_1e_2}^{rs,+}(k_1,k_2,p_0) &\equiv&
-\left(\frac{n_{1,r}(1-n_{2,s})}{p_0+\e_{1,r}+\e_{2,s}}-\frac{(1-n_{1,r})n_{2,s}}
{p_0-\e_{1,r}-\e_{2,s}}\right)(1-N_{1,r}-N_{2,s}) \nonumber \\
&&-\left(\frac{(1-n_{1,r})(1-n_{2,s})}{p_0-\e_{1,r}+\e_{2,s}}-\frac{n_{1,r}
 n_{2,s}}
{p_0+\e_{1,r}-\e_{2,s}}\right)(N_{1,r}-N_{2,s}) \,\, ,  \\
v_{e_1e_2}^{rs,-}(k_1,k_2,p_0) &\equiv&
-\left(\frac{(1-n_{1,r})n_{2,s}}{p_0+\e_{1,r}+\e_{2,s}}-\frac{n_{1,r}(1-n_{2,s})}
{p_0-\e_{1,r}-\e_{2,s}}\right)(1-N_{1,r}-N_{2,s}) \nonumber \\
&&-\left(\frac{n_{1,r} n_{2,s}}{p_0-\e_{1,r}+\e_{2,s}}-\frac{(1-n_{1,r})(1-n_{2,s})}
{p_0+\e_{1,r}-\e_{2,s}}\right)(N_{1,r}-N_{2,s}) \,\, .
\eea
\end{subequations}
Here, we abbreviated
\be \label{abbreviations}
\e_{i,r}\equiv \e^{e_i}_{k_i,r} \quad,\qquad
n_{i,r}\equiv\frac{\e_{i,r}+\m-e_ik_i}{2\e_{i,r}}
\quad,\qquad N_{i,r}\equiv\frac{1}{\exp(\e_{i,r}/T)+1} \qquad
(i=1,2) \,\, .
\ee 
Note that, for $p_0=0$, we have
\be \label{spurious}
v_{e_1e_2}^{rs,+}(k_1,k_2,0)=
v_{e_1e_2}^{rs,-}(k_1,k_2,0)\equiv 
v_{e_1e_2}^{rs}(k_1,k_2,0) \,\, .
\ee

Next we discuss the traces containing the anomalous propagators. Again
the Matsubara sum is completely analogous to the calculation in 
Ref.\ \cite{meissner2}.
Therefore, using Eq.\ (\ref{aprop}), we obtain 
\begin{subequations} \label{matsub2}
\bea 
T\, \sum_{k_0}{\rm Tr}\left[\Gamma_a^\m\Xi^-(K_1)\,\overline{\Gamma}_b^\n\, 
\Xi^+(K_2)\right]&=&\sum_{e_1,e_2}\sum_{r,s}
{\rm Tr}\left[\Gamma_a^\m {\cal M}_{{\bf k}_1}^\dag
{\cal P}_{{\bf k}_1}^r\Lambda_{{\bf k}_1}^{e_1}\overline{\Gamma}_b^\n
\g_0 {\cal M}_{{\bf k}_2} \g_0
{\cal P}_{{\bf k}_2}^s\Lambda_{{\bf k}_2}^{-e_2}\right]\, 
w_{e_1e_2}^{rs}(k_1,k_2,p_0) \,\, , \\
T\, \sum_{k_0}{\rm Tr}\left[\overline{\Gamma}_a^\m\Xi^+(K_1)\,\Gamma_b^\n\, 
\Xi^-(K_2)\right]&=&\sum_{e_1,e_2}\sum_{r,s}
{\rm Tr}\left[\overline{\Gamma}_a^\m \g_0 {\cal M}_{{\bf k}_1} \g_0
{\cal P}_{{\bf k}_1}^r\Lambda_{{\bf k}_1}^{-e_1}\Gamma_b^\n
{\cal M}_{{\bf k}_2}^\dag
{\cal P}_{{\bf k}_2}^s\Lambda_{{\bf k}_2}^{e_2}\right]\, 
w_{e_1e_2}^{rs}(k_1,k_2,p_0) \,\, ,
\eea
\end{subequations}
where (cf.\ Eq.\ (93) of Ref.\ \cite{meissner2})
\bea  \label{defw}
w_{e_1e_2}^{rs}(k_1,k_2,p_0)&\equiv&
\frac{\phi_{1,r}\phi_{2,s}}{4\e_{1,r}\e_{2,s}}
\left[\left(\frac{1}{p_0+\e_{1,r}+\e_{2,s}}-\frac{1}{p_0-\e_{1,r}-\e_{2,s}}\right)
(1-N_{1,r}-N_{2,s}) \right. \nonumber \\
&&\left.-\left(\frac{1}{p_0-\e_{1,r}+\e_{2,s}}-\frac{1}{p_0+\e_{1,r}-\e_{2,s}}
\right)(N_{1,r}-N_{2,s})\right] \,\,. 
\eea
Here,
\be
\phi_{i,r}\equiv\phi^{e_i}(\e_{i,r},{\bf k}_i)
\ee
is the gap function on the quasiparticle mass shell given by the 
excitation branch $k_0=\e_{i,r}$. 
In the derivation of Eqs.\ (\ref{matsub1}) and (\ref{matsub2}), we 
assumed that, in the 2SC and CFL phases, the chemical potentials 
for all $N_f$ quark flavors are the 
same, $\m_1=\ldots=\m_{N_f}\equiv\m$. In this case, the functions $v$ and $w$
only depend on the single chemical potential $\m$ and can thus be factored
out of the flavor trace. In the cases where quarks of the 
same flavor form Cooper pairs, i.e., in the polar and CSL phases,
our formalism also allows for the treatment of a system with $N_f>1$ and
different chemical potentials $\m_n$, $n=1,\ldots,N_f$. 
Then, $v$ and $w$ depend on the quark flavor through $\m_n$ and have to 
be included into the trace over flavor space. 
 
Inserting Eqs.\ (\ref{matsub1}) and
(\ref{matsub2}) into Eq.\ (\ref{nambu}), we obtain for the
general polarization tensor
\bea \label{general}
\Pi_{ab}^{\m\n}(P)&=&\frac{1}{2}\int\frac{d^3\bf k}{(2\p)^3} 
\sum_{e_1,e_2}\sum_{r,s} \Big\{
{\rm Tr}\left[\Gamma_a^\m\,\g_0\,
{\cal P}_{{\bf k}_1}^r\,\Lambda_{{\bf k}_1}^{-e_1} \,  
\Gamma_b^\n \,\g_0\,{\cal P}_{{\bf k}_2}^s\,
\Lambda_{{\bf k}_2}^{-e_2}\right]\, v_{e_1e_2}^{rs,+}
(k_1,k_2,p_0) \nonumber \\
&& \hspace*{2.9cm}+{\rm Tr}\left[\overline{\Gamma}_a^\m\,\g_0\,
{\cal P}_{{\bf k}_1}^r\,\Lambda_{{\bf k}_1}^{e_1} \,  
\overline{\Gamma}_b^\n \,\g_0\,{\cal P}_{{\bf k}_2}^s\,
\Lambda_{{\bf k}_2}^{e_2}\right]\, v_{e_1e_2}^{rs,-}
(k_1,k_2,p_0)  \nonumber \\
&& \hspace*{2.9cm}+{\rm Tr}\left[\Gamma_a^\m{\cal M}_{{\bf k}_1}^\dag
{\cal P}_{{\bf k}_1}^r\Lambda_{{\bf k}_1}^{e_1}\overline{\Gamma}_b^\n
\g_0{\cal M}_{{\bf k}_2}\g_0
{\cal P}_{{\bf k}_2}^s\Lambda_{{\bf k}_2}^{-e_2}\right]\, 
w_{e_1e_2}^{rs}(k_1,k_2,p_0)  \nonumber \\
&&\hspace*{2.9cm}+{\rm Tr}\left[\overline{\Gamma}_a^\m 
\g_0{\cal M}_{{\bf k}_1}\g_0
{\cal P}_{{\bf k}_1}^r\Lambda_{{\bf k}_1}^{-e_1}\Gamma_b^\n
{\cal M}_{{\bf k}_2}^\dag
{\cal P}_{{\bf k}_2}^s\Lambda_{{\bf k}_2}^{e_2}\right]\, 
w_{e_1e_2}^{rs}(k_1,k_2,p_0) \Big\} \,\, .
\eea
In the following we focus on the special case $p_0=0$ and $p\to 0$,
i.e, ${\bf k}_2\to{\bf k}_1\equiv {\bf k}$.  
In this limit, the traces only depend on $\uk$
and the functions $v$ and $w$ only on $k\equiv |{\bf k}|$.  
Thus, the $d^3{\bf k}$ integral factorizes into an angular 
and a radial part. With the abbreviations
\begin{subequations} \label{VWdef}
\bea 
{\cal V}_{ab,e_1e_2}^{\m\n,rs}
&\equiv&\frac{1}{2}\int\frac{d\Omega_{\bf k}}{(2\p)^3}
\left\{ {\rm Tr}\left[\Gamma_a^\m\,\g_0\,{\cal P}_{\bf k}^r\,
\Lambda_{\bf k}^{-e_1}
\,\Gamma_b^\n\,\g_0\,{\cal P}_{\bf k}^s\,\Lambda_{\bf k}^{-e_2}\right] +
{\rm Tr}\left[\overline{\Gamma}_a^\m\,\g_0\,{\cal P}_{\bf k}^r\,
\Lambda_{\bf k}^{e_1}
\,\overline{\Gamma}_b^\n\,\g_0\,{\cal P}_{\bf k}^s\,\Lambda_{\bf k}^{e_2}
\right] \right\} \,\, , 
\\ 
{\cal W}_{ab,e_1 e_2}^{\m\n,rs}&\equiv&
\frac{1}{2}\int\frac{d\Omega_{\bf k}}{(2\p)^3}
\left\{
{\rm Tr}\left[\Gamma_a^\m\,{\cal M}_{\bf k}^\dag\,
{\cal P}_{\bf k}^r
\,\Lambda_{\bf k}^{e_1}\,\overline{\Gamma}_b^\n\,\g_0{\cal M}_{\bf k}\g_0
\,{\cal P}_{\bf k}^s\,\Lambda_{\bf k}^{-e_2}\right] \right.\nonumber \\  
&& \hspace{1.5cm} \left.+ \; {\rm Tr}\left[\overline{\Gamma}_a^\m\,
\g_0{\cal M}_{\bf k}\g_0\,{\cal P}_{\bf k}^r
\,\Lambda_{\bf k}^{-e_1}\,\Gamma_b^\n\,{\cal M}_{\bf k}^\dag
\,{\cal P}_{\bf k}^s\,\Lambda_{\bf k}^{e_2}\right] \right\} \,\, ,
\eea
\end{subequations}
we can write the polarization tensor as
\be \label{polshort}
\Pi_{ab}^{\m\n}(0) \equiv \lim_{p\to 0} \Pi_{ab}^{\m\n}(0,{\bf p})=
\sum_{e_1,e_2}
\sum_{r,s}\left[{\cal V}_{ab,e_1e_2}^{\m\n,rs}
\, \lim\limits_{p\to 0}\int dk\,k^2 v_{e_1e_2}^{rs}(k_1,k_2,0) 
+{\cal W}_{ab,e_1e_2}^{\m\n,rs}
\,\lim\limits_{p\to 0}\int dk\,k^2 w_{e_1e_2}^{rs}(k_1,k_2,0)\right] \,\, .
\ee
Note that only the angular integrals over the color, flavor, and Dirac traces, 
defined by ${\cal V}$ and ${\cal W}$, depend on the symmetries
of the order parameter and thus have to be calculated separately for each 
phase. Therefore, we first consider the $dk$ integrals which are the same 
for all cases. 

In order to see how the two different quasiparticle excitations 
branches (labelled by $r$, $s$), 
as well as normal and
anomalous propagation of the respective excitations (represented by the 
functions $v$, $w$) contribute in the final expressions for the polarization 
tensors in the zero-energy, zero-momentum limit, it is convenient 
to define the quantities
\begin{subequations} \label{kintegrals}
\bea
v^{rs}&\equiv& \frac{1}{\m^2}\lim\limits_{p\to 0}\int dk\,k^2 \left[
v_{++}^{rs}(k_1,k_2,0) + v_{--}^{rs}(k_1,k_2,0)\right] \,\, , \\
\bar{v}^{rs}&\equiv& \frac{1}{\m^2}\lim\limits_{p\to 0}\int dk\,k^2 \left[
v_{+-}^{rs}(k_1,k_2,0) + v_{-+}^{rs}(k_1,k_2,0)\right] \,\, , \\
w^{rs}&\equiv& \frac{1}{\m^2}\lim\limits_{p\to 0}\int dk\,k^2 \left[
w_{++}^{rs}(k_1,k_2,0) + w_{--}^{rs}(k_1,k_2,0)\right] \,\, , \\ 
\bar{w}^{rs}&\equiv& \frac{1}{\m^2}\lim\limits_{p\to 0}\int dk\,k^2 \left[
w_{+-}^{rs}(k_1,k_2,0) + w_{-+}^{rs}(k_1,k_2,0)\right] \,\, .
\eea
\end{subequations}
These quantities are dimensionless since 
$v_{e_1e_2}^{rs}(k_1,k_2,0)$ and $w_{e_1e_2}^{rs}(k_1,k_2,0)$
have the dimension [1/energy] (cf.\ the definitions in Eqs.\ (\ref{defv}) and
(\ref{defw})).
Combining particle-particle ($e_1=e_2=+$) and antiparticle-antiparticle
($e_1=e_2=-$), as well as particle-antiparticle ($e_1=-e_2=\pm$) 
contributions is possible since the corresponding integrals ${\cal V}, 
{\cal W}$ multiplying these
terms turn out to be the same.
In the definitions of $v^{rs}$, $\bar{v}^{rs}$
and $w^{rs}$, $\bar{w}^{rs}$ we divided by the square of the
quark chemical potential $\m^2$ in order to make these quantities 
independent of the quark flavor. This will be 
convenient for the results of the spin-one phases, 
where the formation of Cooper pairs 
is possible for different chemical potentials for each quark flavor.   
In App.\ \ref{AppA} we present the calculation of the relevant 
integrals defined in Eqs.\ (\ref{kintegrals}).
As in Refs.\ \cite{meissner2,meissner3}, we neglect the
antiparticle gap, $\phi^-\simeq 0$, and compute the integrals up to
leading order. In Table \ref{tablevw} we collect all results.
Some integrals vanish since they are proportional to a
vanishing gap. This is the case for $\bar{w}^{rs}=0$ (for all $r,s$), 
since these integrals are proportional to at least one antiparticle gap. 
Clearly, for $T\ge T_c$, the gap vanishes in all phases and thus 
$w^{rs}=\bar{w}^{rs}=0$.   

\begin{table} 
\begin{tabular}{|c|c||c|c|c|c|c|c|c|c||c|c|c|c|c|c|c|c|}
\hline
 & & $v^{11}$ &$v^{22}$ &$v^{12}$ & $v^{21}$ 
&$\bar{v}^{11}$ &$\bar{v}^{22}$ &$\bar{v}^{12}$ & $\bar{v}^{21}$
& $w^{11}$ &$w^{22}$ &$w^{12}$ & $w^{21}$ 
&$\bar{w}^{11}$ &$\bar{w}^{22}$ &$\bar{w}^{12}$ & $\bar{w}^{21}$ \\
\hline
\hline
$T=0$ & 2SC, polar & $-\frac{1}{2}$ & $-1$ & 
\multicolumn{2}{c}{$-\frac{1}{2}$}\vline &  
\multicolumn{4}{c}{$\frac{1}{2}$}\vline\,\vline & $\frac{1}{2}$ & 
\multicolumn{3}{c}{-}\vline &  
0 & \multicolumn{3}{c}{-}\vline \\ 
\cline{2-18}
& CFL, CSL & \multicolumn{4}{c}{$-\frac{1}{2}$}\vline &
\multicolumn{4}{c}{$\frac{1}{2}$}\vline\,\vline & $\frac{1}{8}$ & 
$\frac{1}{2}$ & \multicolumn{2}{c}{$\frac{1}{3}\ln 2$}\vline
 & \multicolumn{4}{c}{0}\vline \\
\hline
$T\ge T_c$ & all phases & \multicolumn{4}{c}{$-1$}\vline &  
\multicolumn{4}{c}{$\frac{1}{2}$}\vline\,\vline & \multicolumn{8}{c}{0} 
\vline\\
\hline
\end{tabular}
\caption{Leading order results  
for the integrals defined in Eqs.\ (\ref{kintegrals}).  
The indices 1 and 2 correspond to the two gaps of each phase (the second
one vanishing in the 2SC and the polar phase), while $v$ 
corresponds to the trace over quasiparticle propagators $G^\pm$ and
$w$ to the trace over anomalous propagators $\Xi^\pm$. The fields with no
entry indicate that these values do not occur in the calculations.}
\label{tablevw}
\end{table} 

In order to discuss the traces and the angular integral in 
Eq.\ (\ref{polshort}), we have to distinguish between the several
phases since the special form of the gap matrix, Eq.\ (\ref{gapmatrix}),
is explicitly involved. Therefore, in the following sections we discuss
the 2SC, CFL, polar, and CSL phases separately
and compute the Debye and Meissner masses for photons and gluons in each 
phase. 
For the 2SC phase \cite{meissner2} and the CFL phase \cite{son,meissner3},
the results for the gluons are already known. Also the masses of
the rotated gauge bosons, where the rotation is given by the new 
generator $\tilde{Q}$, cf.\ Eq.\ (\ref{Qtilde}), were considered for these
two phases \cite{litim}. Nevertheless, we will briefly discuss also these 
cases, since we first want to establish our notation and second, we will show
that for the 2SC phase, there is actually no mixing between electric
gluons and photons, i.e., the longitudinal mixing angle $\theta_2$ is 
zero. Consequently, there is no rotated electric photon.

The basic symmetry properties of all considered phases have been
discussed in detail in Ref.\ \cite{schmitt}. We show
the relevant color-flavor-Dirac matrices in Table \ref{tableMP}.
\begin{table} 
\begin{tabular}{|c||c|c|c|} 
\hline
 & ${\cal M}_{\bf k}$ & ${\cal P}^1_{\bf k}$ & ${\cal P}^2_{\bf k}$ \\
\hline\hline
2SC & $J_3\t_2\g_5$ & $J_3^2$ & $1-J_3^2$ 
\\
\hline
CFL & ${\bf J}\cdot{\bf I}\,\g_5$ & $\frac{1}{3}[({\bf J}\cdot{\bf I})^2
-1)]$ & $\frac{1}{3}[4-({\bf J}\cdot{\bf I})^2]$ \\
\hline
polar &  $\quad J_3[\hat{k}^z+\g_\perp^z(\uk)]\quad$ & 
 $J_3^2$ & $1-J_3^2$ \\
\hline
CSL & $\quad{\bf J}\cdot[\uk+\gperp(\uk)]\quad$ & 
$\quad\frac{1}{3}[\uk+\gperp(\uk)][\uk-\gperp(\uk)]\quad$ & 
$1-{\cal P}^1_{\bf k}$ \\
\hline
\end{tabular}
\caption{Relevant color-flavor-Dirac matrices for the calculation of
the Debye and Meissner masses in a given color-superconducting phase.
The matrix ${\cal M}_{\bf k}$ reflects the symmetries of the 
various gap matrices. For the definition of the projectors 
${\cal P}_{\bf k}^{1,2}$ see Eq.\ (\ref{proj}).
In color space, we use the matrices $(J_i)_{jk} = -i\e_{ijk}$, 
($i,j,k=1,2,3$); in flavor space, we use $(I_i)_{jk} = -i\e_{ijk}$ 
and the second Pauli matrix $\t_2$. In Dirac space, we defined 
$\gperp(\uk)\equiv\vg-\uk\,\vg\cdot\uk$. }\label{tableMP}
\end{table}

\subsection{The 2SC phase} \label{2SCphase}

\subsubsection{Gluon polarization tensor ($a,b\le 8$)}

Inserting the matrices given in the second line of Table \ref{tableMP} into 
Eqs.\ (\ref{VWdef}), we obtain 
\begin{subequations} \label{VW2SCgluon}
\bea
{\cal V}_{ab,e_1e_2}^{\m\n,rs}&=&
g^2\left\{{\rm Tr}[T_a\,{\cal P}^r\,T_b\,{\cal P}^s]
\;{\cal T}_{-e_1,-e_2}^{\m\n}
 + {\rm Tr}[T_a^T\,{\cal P}^r\,T_b^T\,{\cal P}^s]\;{\cal T}_{e_1,e_2}^{\m\n}
\right\} \,\, ,
 \\
{\cal W}_{ab,e_1e_2}^{\m\n,rs}&=&g^2\left\{{\rm Tr}[T_a\,J_3\,{\cal P}^r\,T_b^T
\,J_3\,{\cal P}^s]\;{\cal U}_{e_1,-e_2}^{\m\n} + 
{\rm Tr}[T_a^T\,J_3\,{\cal P}^r\,T_b\,J_3\,{\cal P}^s]
\;{\cal U}_{-e_1,e_2}^{\m\n}\right\} \,\, , 
\eea
\end{subequations}
where the traces only run over color space and where we defined
\begin{subequations}\label{TUdef} 
\bea 
{\cal T}_{e_1,e_2}^{\m\n}&\equiv&\int\frac{d\Omega_{\bf k}}{(2\p)^3}
{\rm Tr}\left[\g^\m\,\g_0\,\Lambda_{\bf k}^{e_1}
\,\g^\n\,\g_0\,\Lambda_{\bf k}^{e_2}\right] \,\, , \\
{\cal U}_{e_1,e_2}^{\m\n}&\equiv&\int\frac{d\Omega_{\bf k}}{(2\p)^3}
{\rm Tr}\left[\g^\m\,\g^5\,\Lambda_{\bf k}^{e_1}\,\g^\n
\,\g^5\,\Lambda_{\bf k}^{e_2}\right] \,\, .
\eea
\end{subequations}
Here, the traces only run over Dirac space.
We used the fact that the projectors 
${\cal P}^{r,s}\equiv{\cal P}_{\bf k}^{r,s}$
do not depend on the quark momentum ${\bf k}$ and that 
the color and Dirac traces factorize. 
Furthermore, the trivial flavor trace was already performed, yielding a 
factor 2. The angular integrals over Dirac traces, Eqs. (\ref{TUdef}), are
easily evaluated,
\begin{subequations} \label{TU}
\bea
{\cal T}^{00}_{e_1,e_2}&=&-{\cal U}^{00}_{e_1,-e_2}=\frac{1}{2\p^2}
(1+e_1e_2) 
\,\, , \label{TU00}\\
{\cal T}^{0i}_{e_1,e_2}&=&{\cal T}^{i0}_{e_1,e_2}=
{\cal U}^{0i}_{e_1,-e_2}={\cal U}^{i0}_{e_1,-e_2}=0 
\,\, , \label{TU0i}\\
{\cal T}^{ij}_{e_1,e_2}&=&{\cal U}^{ij}_{e_1,-e_2}=\frac{1}{2\p^2}\d_{ij}
(1-\frac{1}{3}e_1e_2)     \,\, , \label{TUij}
\eea
\end{subequations}
where $i,j=1,2,3$. 
For the evaluation of the color traces note that
$J_3{\cal P}^1=J_3$ and $J_3{\cal P}^2=0$. 
We find that ${\cal V}$ and ${\cal W}$, given in Eqs.\ (\ref{VW2SCgluon}),
are diagonal in the adjoint color indices $a$ and $b$.

\begin{enumerate}
\renewcommand{\labelenumi}{(\alph{enumi})}
\item $\m=\n=0$.
With Eq.\ (\ref{TU00}) we obtain after performing the color traces
\be \label{2SCgluon00}
\Pi^{00}_{ab}(0)=\d_{ab}\frac{2g^2\m^2}{\p^2}\left\{
\begin{array}{cl} \frac{1}{2}v^{11}+\frac{1}{2}w^{11} &
\quad \mbox{for} \quad a=1,2,3 \,\, , \\ \\
\frac{1}{4}(v^{12}+v^{21}) & \quad \mbox{for} \quad 
a=4-7 \,\, ,\\ \\
\frac{1}{6}v^{11}+\frac{1}{3}v^{22}-\frac{1}{6}w^{11} &
\quad \mbox{for} \quad a=8 \,\, . 
\end{array}
\right. 
\ee 

\item $\m=0,\n=i$ and $\m=i,\n=0$.
Due to Eq.\ (\ref{TU0i}), 
\be
\Pi_{ab}^{0i}(0) = \Pi_{ab}^{i0}(0) = 0 \,\, .
\ee 

\item $\m=i,\n=j$.
Due to Eq.\ (\ref{TUij}), the polarization tensor is diagonal
in spatial indices $i$, $j$. We obtain 
\be \label{2SCgluonij}
\Pi^{ij}_{ab}(0)=\d_{ab}\d^{ij}\frac{2g^2\m^2}{3\p^2}
\left\{
\begin{array}{cl} \frac{1}{2}(v^{11}+2\bar{v}^{11}) - 
\frac{1}{2}(w^{11}+2\bar{w}^{11}) &
\quad \mbox{for} \quad a=1,2,3 \,\, ,\\ \\
\frac{1}{4}[(v^{12}+v^{21})+2(\bar{v}^{12}+\bar{v}^{21})]
 & \quad \mbox{for} \quad 
a=4-7 \,\, ,\\ \\
\frac{1}{6}(v^{11}+2\bar{v}^{11})+\frac{1}{3}(v^{22}
+2\bar{v}^{22})+\frac{1}{6}(w^{11}+2\bar{w}^{11}) &
\quad \mbox{for} \quad a=8 \,\, . 
\end{array}
\right. 
\ee

\end{enumerate}

\subsubsection{Mixed polarization tensor ($a\le 8, b=9$ and $a=9, b\le 8$)}

For a system in the color-superconducting 2SC phase where quarks
with the electric charges $q_1$ and $q_2$ form Cooper pairs,
the electric charge
generator, introduced in Eq.\ (\ref{quarkpartition}), is given by
$Q={\rm diag}(q_1,q_2)$. Thus, we obtain
\begin{subequations} \label{VW2SCmixed}
\bea
{\cal V}_{a\g,e_1e_2}^{\m\n,rs}&=& 
\frac{1}{2}\,eg(q_1+q_2)\left\{ {\rm Tr}[T_a\,{\cal P}^r\,
{\cal P}^s]\,{\cal T}_{-e_1,-e_2}^{\m\n} +{\rm Tr}[T_a^T\,{\cal P}^r\,
{\cal P}^s]\,{\cal T}_{e_1,e_2}^{\m\n} \right\} \,\, , \\
{\cal W}_{a\g,e_1e_2}^{\m\n,rs}&=&
\frac{1}{2}\,eg(q_1+q_2)\left\{ {\rm Tr}[T_a\,J_3\,{\cal P}^r\,
J_3\,{\cal P}^s]\,
{\cal U}_{e_1,-e_2}^{\m\n} + {\rm Tr}[T_a^T\,J_3\,{\cal P}^r\,J_3\,
{\cal P}^s]\,{\cal U}_{-e_1,e_2}^{\m\n}\right\}
\,\, . 
\eea
\end{subequations}
It is not difficult to show that the polarization tensor 
is symmetric under the exchange of photon and gluon indices, 
\be \label{symmetric}
\Pi_{a\g}^{\m\n}(0)=
\Pi_{\g a}^{\m\n}(0) \,\, .
\ee
Using $J_3^3=J_3$, all color traces in Eq.\ (\ref{VW2SCmixed}) reduce 
to ${\rm Tr}[T_a J_3^2]=\d_{a8}/\sqrt{3}$. Therefore, we obtain
for the various Dirac components of the tensor:

\begin{enumerate}
\renewcommand{\labelenumi}{(\alph{enumi})}
\item $\m=\n=0$.
\be \label{2SCmixed00}
\Pi_{a\g}^{00}(0)=(q_1+q_2)\frac{eg\,\m^2}{\sqrt{3}\p^2}\left\{
\begin{array}{cl} 0 &  \quad \mbox{for} \quad a=1-7 \,\, ,\\ \\
v^{11}-v^{22}-w^{11} &  \quad \mbox{for} \quad a=8 \,\, .\\
\end{array} \right.
\ee

\item $\m=0,\n=i$ and $\m=i,\n=0$.
\be
\Pi_{a\g}^{0i}(0) = \Pi_{a\g}^{i0}(0) = 0 \,\, .
\ee 

\item $\m=i,\n=j$.
\be \label{2SCmixedij}
\Pi_{a\g}^{ij}(0)=\d_{ij}(q_1+q_2)\frac{eg\,\m^2}{3\sqrt{3}\p^2}\left\{
\begin{array}{cl} 0 &  \quad \mbox{for} \quad a=1-7 \,\, ,\\ \\
(v^{11}+2\bar{v}^{11})-(v^{22}+2\bar{v}^{22})+
(w^{11}+2\bar{w}^{11}) &  \quad \mbox{for} \quad a=8 \,\, .\\
\end{array} \right.
\ee

\end{enumerate}

\subsubsection{Photon polarization tensor ($a=b=9$)}

In this case, the tensors ${\cal V}$, ${\cal W}$ are
\begin{subequations} \label{VW2SCphoton}
\bea
{\cal V}_{\g\g,e_1e_2}^{\m\n,rs}&=& 
\frac{1}{2}\,e^2(q_1^2+q_2^2)\,{\rm Tr}[{\cal P}^r\,
{\cal P}^s]\,\left( {\cal T}_{-e_1,-e_2}^{\m\n} + {\cal T}_{e_1,e_2}^{\m\n}
\right) \,\, ,
\\
{\cal W}_{\g\g,e_1e_2}^{\m\n,rs}&=&
e^2q_1q_2\,{\rm Tr}[J_3\,{\cal P}^r\,J_3\,{\cal P}^s]
\,\left( {\cal U}_{e_1,-e_2}^{\m\n} +{\cal U}_{-e_1,e_2}^{\m\n} \right) 
\,\, . 
\eea
\end{subequations}
Here, we performed the flavor traces ${\rm Tr}[Q^2]=q_1^2+q_2^2$ and 
${\rm Tr}[Q\t_2Q\t_2]=2q_1q_2$.
After performing the color traces and the sums over $e_1$, $e_2$
and $r$, $s$, the results for the different components are as follows. 

\begin{enumerate}
\renewcommand{\labelenumi}{(\alph{enumi})}
\item $\m=\n=0$.
\be \label{2SCphoton00}
\Pi_{\g\g}^{00}(0)=\frac{e^2\m^2}{\p^2}
\left[(q_1^2+q_2^2)\,(2v^{11}+v^{22})
-4q_1q_2w^{11}\right]
\,\, .
\ee

\item $\m=0,\n=i$ and $\m=i,\n=0$.
\be
\Pi_{\g\g}^{0i}(0) = \Pi_{\g\g}^{i0}(0) = 0 \,\, .
\ee 

\item $\m=i,\n=j$.
\be  \label{2SCphotonij}
\Pi_{\g\g}^{ij}(0)=\d_{ij}
\frac{e^2\m^2}{3\p^2}
\left\{(q_1^2+q_2^2)\,\left[2(v^{11}+2\bar{v}^{11})+
(v^{22}+2\bar{v}^{22})\right]+
4q_1q_2\,(w^{11}+2\bar{w}^{11})\right\}  \,\, .
\ee

\end{enumerate}

\subsection{The CFL phase} \label{CFLphase}

\subsubsection{Gluon polarization tensor ($a,b\le 8$)}

With the matrix ${\cal M}_{\bf k}$ and the projectors 
${\cal P}_{\bf k}^{1,2}$ for the CFL phase, given in Table \ref{tableMP},
Eqs.\ (\ref{VWdef}) become
\begin{subequations} \label{VWCFLgluon}
\bea
{\cal V}_{ab,e_1e_2}^{\m\n,rs}&=&\frac{1}{2}\,g^2\left\{ 
{\rm Tr}[T_a\,{\cal P}^r\,T_b\,{\cal P}^s]\,{\cal T}_{-e_1,-e_2}^{\m\n} +  
{\rm Tr}[T_a^T\,{\cal P}^r\,T_b^T\,{\cal P}^s]\,{\cal T}_{e_1 e_2}^{\m\n}
\right\} \,\, ,
\\
{\cal W}_{ab,e_1e_2}^{\m\n,rs}&=&
\frac{1}{2}\,g^2\left\{ {\rm Tr}\left[T_a\,{\bf J}\cdot{\bf I}
\,{\cal P}^r\,T_b^T\,{\bf J}\cdot{\bf I}\,{\cal P}^s\right]
\,{\cal U}_{e_1,-e_2}^{\m\n} + {\rm Tr}\left[T_a^T\,{\bf J}\cdot{\bf I}
\,{\cal P}^r\,T_b\,{\bf J}\cdot{\bf I}\,{\cal P}^s\right]
\,{\cal U}_{-e_1,e_2}^{\m\n} \right\} \,\, . 
\eea
\end{subequations}
As in the 2SC phase, the projectors ${\cal P}^r$, ${\cal P}^s$ do not
depend on momentum. Consequently, the angular integrals defined in 
Eq.\ (\ref{TUdef}) also appear in the CFL phase. Therefore,
also in the CFL phase the $(0i)$ and $(i0)$ components of the
polarization tensor vanish, and the $(ij)$ components are proportional
to $\d^{ij}$. Contrary to the 2SC
phase, color and flavor traces cannot be performed separately, since the
projectors ${\cal P}^r$, ${\cal P}^s$ are nontrivial matrices both
in color and in flavor space. 
In order to perform the color-flavor trace,
one uses the relations ${\rm Tr}[T_a{\cal P}^1T_b{\cal P}^1]=0$ and 
${\bf J}\cdot{\bf I}\,{\cal P}^1=-2\,{\cal P}^1$.  
The polarization tensor is not only diagonal in color, but also 
has the same value for all eight gluons. One finally obtains the 
following expressions for the gluon polarization tensors.
\begin{enumerate}
\renewcommand{\labelenumi}{(\alph{enumi})}
\item $\m=\n=0$.

\be \label{CFLgluon00}
\Pi_{ab}^{00}(0)=\d_{ab}\frac{g^2\m^2}{6\p^2}
\left[(v^{12}+v^{21})+7v^{22}+2(w^{12}+w^{21})
+2w^{22}\right] \,\, .
\ee

\item $\m=0,\n=i$ and $\m=i,\n=0$.
\be
\Pi_{ab}^{0i}(0) = \Pi_{ab}^{i0}(0) = 0 \,\, .
\ee 

\item $\m=i,\n=j$.
\bea \label{CFLgluonij}
\Pi_{ab}^{ij}(0)&=&\d_{ij}\d_{ab}\frac{g^2\m^2}{18\p^2}
\left[(v^{12}+v^{21})+2(\bar{v}^{12}+\bar{v}^{21})+
7(v^{22}+2\bar{v}^{22}) \right.\nonumber \\
&&\left. \hspace{1.7cm} -2(w^{12}+w^{21})
-4(\bar{w}^{12}+\bar{w}^{21})
-2(w^{22}+2\bar{w}^{22})\right] \,\, .
\eea
\end{enumerate}

\subsubsection{Mixed polarization tensor ($a\le 8, b=9$ and $a=9, b\le 8$)}

To compute the mixed polarization tensors, we need the electric charge
generator $Q$. Since we consider a system with three quark flavors of electric 
charges $q_1$, $q_2$, $q_3$, we have $Q={\rm diag}(q_1,q_2,q_3)$. In the final
result we will insert the charges for $u$, $d$, and $s$ quarks.
We obtain
\begin{subequations} \label{VWCFLmixed}
\bea
{\cal V}_{a\g,e_1e_2}^{\m\n,rs}&=&\frac{1}{2}\,eg\,\left\{ 
{\rm Tr}[T_a\,{\cal P}^r\,Q\,{\cal P}^s]\,{\cal T}_{-e_1,-e_2}^{\m\n} + 
{\rm Tr}[T_a^T\,{\cal P}^r\,Q\,{\cal P}^s]\,{\cal T}_{e_1 e_2}^{\m\n} \right\}
 \,\, ,\\
{\cal W}_{a\g,e_1e_2}^{\m\n,rs}&=&
\frac{1}{2}\,eg\,\left\{{\rm Tr}\left[T_a\,{\bf J}\cdot{\bf I}
\,{\cal P}^r\,Q\,{\bf J}\cdot{\bf I}\,{\cal P}^s\right]
\,{\cal U}_{e_1,-e_2}^{\m\n} + {\rm Tr}\left[T_a^T\,{\bf J}\cdot{\bf I}
\,{\cal P}^r\,Q\,{\bf J}\cdot{\bf I}\,{\cal P}^s\right]
\,{\cal U}_{-e_1,e_2}^{\m\n} \right\}\,\, . 
\eea
\end{subequations}
First we note that Eq.\ (\ref{symmetric}) also holds for the CFL phase.
With the help of the relations 
\be
{\rm Tr}[T_a{\cal P}^1Q{\cal P}^1]=0 \quad, \qquad
{\rm Tr}[T_aQ]={\rm Tr}[T_a]{\rm Tr}[Q]=0 \,\, , 
\ee
and 
\be
{\rm Tr}[T_a{\bf J}\cdot{\bf I}\,Q\,{\bf J}\cdot{\bf I}]=
3\,{\rm Tr}[T_a{\cal P}^1Q]=
3\,{\rm Tr}[T_a{\cal P}^1Q\,{\bf J}\cdot{\bf I}]=\d_{a3}\frac{1}{2}(q_1-q_2)
+\d_{a8}\frac{1}{2\sqrt{3}}(q_1+q_2-2q_3) 
\ee
we obtain the following results.

\begin{enumerate}
\renewcommand{\labelenumi}{(\alph{enumi})}
\item $\m=\n=0$.
\be \label{CFLmixed00}
\Pi_{a\g}^{00}(0)=\frac{eg\,\m^2}{6\p^2}\left\{
\begin{array}{cl} 0 &  \quad \mbox{for} \quad a=1,2,4-7 \,\, ,\\ \\
(q_1-q_2)
\left[(v^{12}+v^{21})-2v^{22}+2(w^{12}+
w^{21})-7w^{22})\right] &  \quad \mbox{for} \quad a=3 \,\, ,\\ \\
\frac{1}{\sqrt{3}}(q_1+q_2-2q_3)
\left[(v^{12}+v^{21})-2v^{22}+2(w^{12}+
w^{21})-7w^{22})\right] &  \quad \mbox{for} \quad a=8 \,\, .\\
\end{array} \right.
\ee

\item $\m=0,\n=i$ and $\m=i,\n=0$.
\be
\Pi_{a\g}^{0i}(0) = \Pi_{a\g}^{i0}(0) = 0 \,\, .
\ee 

\item $\m=i,\n=j$.
\be \label{CFLmixedij}
\Pi_{a\g}^{ij}(0)=\d^{ij}\frac{eg\,\m^2}{18\p^2}\left\{
\begin{array}{cl} 0 &  \quad \mbox{for} \quad a=1,2,4-7 \,\, ,\\ \\
(q_1-q_2)
\left[(v^{12}+v^{21})+2(\bar{v}^{12}+\bar{v}^{21})
-2v^{22}-4\bar{v}^{22} \right. & \\
\left.-2(w^{12}+w^{21})-4(\bar{w}^{12}+\bar{w}^{21})
+7w^{22}+14\bar{w}^{22}\right] &  \quad \mbox{for} \quad a=3 \,\, ,\\ \\
\frac{1}{\sqrt{3}}(q_1+q_2-2q_3)
\left[(v^{12}+v^{21})+2(\bar{v}^{12}+\bar{v}^{21})
-2v^{22}-4\bar{v}^{22} \right. & \\
\left.-2(w^{12}+w^{21})-4(\bar{w}^{12}+\bar{w}^{21})
+7w^{22}+14\bar{w}^{22}\right] &  \quad \mbox{for} \quad a=8 \,\, .\\
\end{array} \right.
\ee
\end{enumerate}

\subsubsection{Photon polarization tensor ($a=b=9$)}

In this case, the tensors ${\cal V}$, ${\cal W}$ read
\begin{subequations} \label{VWCFLphoton}
\bea
{\cal V}_{\g\g,e_1e_2}^{\m\n,rs}&=&\frac{1}{2}\,e^2 
{\rm Tr}[Q\,{\cal P}^r\,Q\,{\cal P}^s]\,\left({\cal T}_{-e_1,-e_2}^{\m\n} + 
{\cal T}_{e_1,e_2}^{\m\n} \right) \,\, ,\\
{\cal W}_{\g\g,e_1e_2}^{\m\n,rs}&=&
\frac{1}{2}\,e^2{\rm Tr}\left[Q\,{\bf J}\cdot{\bf I}
\,{\cal P}^r\,Q\,{\bf J}\cdot{\bf I}\,{\cal P}^s\right]
\,\left({\cal U}_{e_1,-e_2}^{\m\n} + {\cal U}_{-e_1,e_2}^{\m\n}\right)\,\, . 
\eea
\end{subequations}
In order to perform the color-flavor traces, we abbreviate
the following sums over the three quark charges,
\be
x\equiv\sum_{n,m=1}^3 q_nq_m \quad, \qquad y\equiv\sum_{n=1}^3 q_n^2 \,\, .
\ee
Then, with ${\rm Tr}[Q\,{\cal P}^1Q\,{\cal P}^1]=x/9$, 
${\rm Tr}[Q^2{\bf 1}_c]=9\,{\rm Tr}[Q\,{\cal P}^1Q]=3y$ (where ${\bf 1}_c$
is the unit matrix in color space), and 
${\rm Tr}[Q\,{\bf J}\cdot{\bf I}\,Q\,{\bf J}\cdot{\bf I}]=
3{\rm Tr}[Q\,{\bf J}\cdot{\bf I}\,{\cal P}^1Q\,{\bf J}\cdot{\bf I}]=2(x-y)$,
we obtain the following results.

\begin{enumerate}
\renewcommand{\labelenumi}{(\alph{enumi})}
\item $\m=\n=0$.
\bea \label{CFLphoton00}
\Pi_{\g\g}^{00}(0)&=&\frac{e^2\m^2}{9\p^2}
\left[x\,v^{11}+(3y-x)(v^{12}+v^{21})+(21y+x)v^{22}
\right. \nonumber \\
&&\left. \hspace{0.8cm}- 4x\,w^{11}+(6y-2x)(w^{12}+w^{21})
+(6y-10x)w^{22}   \right]
\,\, .
\eea
In a system of $d$, $s$, and $u$ quarks we have the electric charges
$q_1=q_2=-1/3$ and $q_3=2/3$. In this case,
\be \label{uds}
x=0 \quad, \qquad y=\frac{2}{3} \,\, .
\ee
Inserting these values into Eq.\ (\ref{CFLphoton00}), the result 
becomes proportional to the gluon polarization tensor, 
Eq.\ (\ref{CFLgluon00}),
\be 
\Pi_{\g\g}^{00}(0)=\frac{4}{3}\frac{e^2}{g^2}\Pi_{aa}^{00}(0) \,\, .
\ee

\item $\m=0,\n=i$ and $\m=i,\n=0$.
\be
\Pi_{\g\g}^{0i}(0) = \Pi_{\g\g}^{i0}(0) = 0 \,\, .
\ee 

\item $\m=i,\n=j$.
\bea \label{CFLphotonij}
\Pi_{\g\g}^{ij}(0)&=&\d_{ij}\frac{e^2\m^2}{27\p^2}
\left\{x\,(v^{11}+2\bar{v}^{11})+(3y-x)\left[v^{12}+v^{21}
+2(\bar{v}^{12}+\bar{v}^{21})\right] \right.\nonumber \\
&&\left.\hspace{1.5cm} +\,(21y+x)(v^{22}+2\bar{v}^{22})
+ 4x\,(w^{11}+2\bar{w}^{11})
\right. \nonumber \\
&&\left. \hspace{1.5cm} -\,(6y-2x)\left[
w^{12}+w^{21}+2(\bar{w}^{12}+\bar{w}^{21})\right]
-(6y-10x)(w^{22}+2\bar{w}^{22})   \right\}
\,\, .
\eea
Again, using the quark charges of $d$, $s$, and $u$ quarks that lead to
Eq.\ (\ref{uds}), we obtain a result proportional to that given in 
Eq.\ (\ref{CFLgluonij}),
\be
\Pi_{\g\g}^{ij}(0)=\frac{4}{3}\frac{e^2}{g^2}
\Pi_{aa}^{ij}(0) \,\, .
\ee

\end{enumerate}

\subsection{The polar phase} \label{polarphase}

In the polar phase, Cooper pairs are formed by quarks of a single flavor
and carry total spin one, $J=1$. This phase is defined by the matrix 
${\cal M}_{\bf k}=J_3[\hat{k}^z+\g_\perp^z(\uk)]$, cf.\ Table \ref{tableMP}.
The spin of the Cooper pair is aligned with the spatial $z$ direction
which means that rotational $SO(3)_J$ is broken to $SO(2)_J$, cf.\ Table
\ref{tablesymmetry}. As in the 2SC phase, the condensate also points in a 
fixed color direction. Physically, this means that quarks of one color, 
say blue, remain unpaired. This similarity to the 2SC phase can also
be seen in the projectors ${\cal P}^{1,2}$ which are identical in both 
phases. Note that ${\cal P}^{1,2}$ in the polar phase do not depend 
on $\uk$ although ${\cal M}_{\bf k}$ does. In the following we consider
a system of quarks with $N_f$ different flavors where each quark flavor
forms Cooper pairs separately. Each quark flavor has a separate electric
charge, $q_1,\ldots,q_{N_f}$, and chemical potential, $\m_1,\ldots,\m_{N_f}$.

\subsubsection{Gluon polarization tensor ($a,b\le 8$)}

In this case, we have
\begin{subequations} \label{VWpolargluon}
\bea
{\cal V}_{ab,e_1e_2}^{\m\n,rs}&=&\frac{1}{2}\,g^2
\left\{{\rm Tr}[T_a\,{\cal P}^r\,T_b\,{\cal P}^s]
\,{\cal T}_{-e_1,-e_2}^{\m\n} + {\rm Tr}[T_a^T\,{\cal P}^r\,T_b^T\,{\cal P}^s]
\,{\cal T}_{e_1,e_2}^{\m\n}\right\} \,\, ,\\
{\cal W}_{ab,e_1e_2}^{\m\n,rs}&=&\frac{1}{2}\,g^2
\left\{{\rm Tr}[T_a\,J_3\,{\cal P}^r\,T_b^T
\,J_3\,{\cal P}^s]\,\hat{{\cal U}}_{e_1,-e_2}^{\m\n} +
{\rm Tr}[T_a^T\,J_3\,{\cal P}^r\,T_b
\,J_3\,{\cal P}^s]\,\hat{{\cal U}}_{-e_1,e_2}^{\m\n} \right\} \,\, , 
\eea
\end{subequations}
where
\be
\hat{{\cal U}}_{e_1,e_2}^{\m\n}\equiv -\int\frac{d\Omega_{\bf k}}{(2\p)^3}
{\rm Tr}[\g^\m\,(\hat{k}^z-\g^z_\perp(\uk))\,\Lambda_{\bf k}^{e_1}
\,\g^\n\,(\hat{k}^z-\g^z_\perp(\uk))\,\Lambda_{\bf k}^{e_2}] \,\, .
\ee
Since 
\be
{\rm Tr}[\g^\m\,(\hat{k}^z-\g^z_\perp(\uk))\,\Lambda_{\bf k}^{e_1}
\,\g^\n\,(\hat{k}^z-\g^z_\perp(\uk))\,\Lambda_{\bf k}^{-e_2}]=
{\rm Tr}[\g^\m\,\Lambda_{\bf k}^{e_1}\,\g^\n\,\Lambda_{\bf k}^{-e_2}] \,\, ,
\ee
we find
\be \label{2SCpolar}
\hat{{\cal U}}_{e_1,e_2}^{\m\n}={\cal U}_{e_1,e_2}^{\m\n} \,\, .
\ee
Consequently, the gluon polarization tensor in
the polar phase is almost identical to that in the
2SC phase. The only difference is the flavor trace
which here yields a factor $\sum_{n=1}^{N_f}\m_n^2$. 
Therefore, Eqs.\ (\ref{2SCgluon00}) -- 
(\ref{2SCgluonij}) hold also for the polar phase 
after replacing $2\m^2$ with $\sum_{n=1}^{N_f}\m_n^2$.

\subsubsection{Mixed polarization tensor ($a\le 8, b=9$ and $a=9, b\le 8$)}

Obviously, also for the mixed polarization tensor, the Dirac and 
color part is identical to the 2SC case. Consequently, in 
Eqs.\ (\ref{2SCmixed00}) -- (\ref{2SCmixedij}) one has to replace 
the factor $(q_1+q_2)\m^2$ by $\sum_{n=1}^{N_f} q_n\m_n^2$ in order to 
obtain the results for the polar phase. 

\subsubsection{Photon polarization tensor ($a=b=9$)}

Analogously, for the photon polarization tensor, the 2SC results
given in Eqs.\ (\ref{2SCphoton00}) -- (\ref{2SCphotonij}) are
valid for the polar phase with the following modifications. 
Since in the 2SC phase there is a nontrivial flavor structure 
in the matrix ${\cal M}_{\bf k}$, there are two different flavor 
factors in Eqs.\ (\ref{2SCphoton00}) -- (\ref{2SCphotonij}), 
namely $(q_1^2+q_2^2)\m^2$ and $2q_1q_2\m^2$. 
Replacing each of these factors
by the common factor $\sum_{n=1}^{N_f}q_n^2\m_n^2$ yields the corresponding
results for the polar phase.

\subsection{The CSL phase} \label{CSLphase}

As in Sec.\ \ref{polarphase}, we consider a system of $N_f$ quark flavors,
each flavor forming Cooper pairs separately. In the CSL phase, all
quark colors acquire two gapped excitation energies. This is
similar to the CFL phase, where there is also a quasiparticle excitation 
branch with an energy gap $\phi$ and a second one with an energy
gap $2\phi$. Formally, this structure has its origin in the two 
nonzero eigenvalues of the operator 
$\g_0{\cal M}_{\bf k}^\dag{\cal M}_{\bf k}\g_0$, cf.\ Eqs.\ (\ref{excite})
and (\ref{proj}). In the case of the CSL phase, the matrix 
${\cal M}_{\bf k}$ as well as the projectors ${\cal P}_{\bf k}^{1,2}$
depend on the direction of the quark momentum $\uk$. Due to 
color-spin locking, the spatial components of the vector $\uk$ are
aligned with the three color directions red, green, and blue.

\subsubsection{Gluon polarization tensor ($a,b\le 8$)}

Inserting the matrices ${\cal M}_{\bf k}$ and ${\cal P}_{\bf k}^{1,2}$
from Table \ref{tableMP} into Eq.\ (\ref{VWdef}) yields   
\begin{subequations} \label{VWCSLgluon}
\bea 
{\cal V}_{ab,e_1e_2}^{\m\n,rs}
&=&\frac{g^2}{2}\,\int\frac{d\Omega_{\bf k}}{(2\p)^3}
\Big\{{\rm Tr}\left[\g^\m\,\g_0\,T_a\,{\cal P}_{\bf k}^r\,
\Lambda_{\bf k}^{-e_1}\,\g^\n\,\g_0\,T_b\,{\cal P}_{\bf k}^s
\Lambda_{\bf k}^{-e_2}\right] + \left(e_{1,2}\to -e_{1,2},\; 
T_{a,b}\to T_{a,b}^T \right) \Big\} \,\, , \\
{\cal W}_{ab,e_1e_2}^{\m\n,rs}
&=&-\frac{g^2}{2}\, \int\frac{d\Omega_{\bf k}}{(2\p)^3}
\Big\{ {\rm Tr}\left[\g^\m\,T_a\,{\bf J}\cdot(\uk-\gperp(\uk))
\,{\cal P}_{\bf k}^r
\,\Lambda_{\bf k}^{e_1}\,\g_\n\,T_b^T\,{\bf J}\cdot(\uk-\gperp(\uk))\,
{\cal P}_{\bf k}^s\,\Lambda_{\bf k}^{-e_2}\right] \nonumber \\
&&\hspace{2.2cm}+ \left(e_{1,2}\to -e_{1,2}, \;
T_{a,b}\to T_{a,b}^T \right) \Big\} \,\, .
\eea
\end{subequations}
The trace over the $12\times 12$ color-Dirac
matrices is more complicated than in all previously discussed cases. 
For the $(00)$ components, both ${\cal V}$ and 
${\cal W}$ are diagonal in color space. The diagonal elements
are collected in Table \ref{tableCSL1}. They are divided into two
parts, one corresponding to the symmetric Gell-Mann matrices,
$a=1,3,4,6,8$, the other corresponding to the antisymmetric Gell-Mann
matrices, $a=2,5,7$. 
\begin{table}
\begin{tabular}{|c||c|c|c|}
\hline
$a$ & ${\cal V}_{aa,e_1e_2}^{00,11}$ & 
${\cal V}_{aa,e_1e_2}^{00,12}={\cal V}_{aa,e_1e_2}^{00,21}$ & 
${\cal V}_{aa,e_1e_2}^{00,22}$ \\
\hline
\hline
$\,\,1,3,4,6,8\,\,$ & 0 & $(1+e_1e_2)/(12\p^2)$ & 
$(1+e_1e_2)/(12\p^2)$ \\
\hline
$2,5,7$ & $\,\,(1+e_1e_2)/(18\p^2)\,\,$ & $\,\,(1+e_1e_2)/(36\p^2)\,\,$ & 
$\,\,5(1+e_1e_2)/(36\p^2)\,\,$ \\
\hline 
\hline
 & ${\cal W}_{aa,e_1e_2}^{00,11}$ & 
${\cal W}_{aa,e_1e_2}^{00,12}={\cal W}_{aa,e_1e_2}^{00,21}$ & 
${\cal W}_{aa,e_1e_2}^{00,22}$ \\
\hline
\hline
$\,\,1,3,4,6,8\,\,$ & 0 & $(1+e_1e_2)/(6\p^2)$ 
 & $-(1+e_1e_2)/(12\p^2)$ \\
\hline
$2,5,7$ & $\,\,2(1+e_1e_2)/(9\p^2)\,\,$ & 
$\,\,-(1+e_1e_2)/(18\p^2)\,\,$ & $\,\,5(1+e_1e_2)/(36\p^2)\,\,$ \\
\hline
\end{tabular}
\caption{$(00)$ components of the tensors ${\cal V}$, ${\cal W}$ for the
gluon polarization tensor in the CSL phase, defined in 
Eqs.\ (\ref{VWCSLgluon}). All tensors are diagonal in color
space.  }
\label{tableCSL1}
\end{table}
For the $(i0)$ and $(0i)$ components, we find (omitting the indices $r$, $s$
and $e_1$, $e_2$)
\be \label{VW0i}
{\cal V}_{ab}^{i0}={\cal V}_{ab}^{0i}=
{\cal W}_{ab}^{i0}={\cal W}_{ab}^{0i}=0 \,\, .
\ee
Therefore, also for the CSL phase, the $(0i)$ and $(i0)$ components 
of the gluon polarization tensor vanish.

For the $(ij)$ components, we find that ${\cal V}$
and ${\cal W}$ are neither diagonal with respect to
color $a$, $b$ nor with respect to spatial indices. 
Nevertheless, after inserting the $k$ integrals from
Table \ref{tablevw}, the gluon polarization tensor becomes diagonal,
i.e., $\Pi_{ab}^{ij}(0)\sim\d_{ab}\d^{ij}$. The reason
for the cancellation of all non-diagonal elements are the following properties
of ${\cal V}$ and ${\cal W}$,
\begin{subequations} \label{VWijgluon}
\bea
\sum_{rs}{\cal V}^{ij,rs}_{ab,e_1e_2}&\sim&\d_{ab}\d^{ij}\frac{1}{6\p^2} 
\label{Vijgluon} \,\, ,\\
{\cal W}_{ab,e_1e_2}^{ij,rs}&\sim& 1-e_1e_2 \qquad \mbox{for ($i\neq j$;
$a$, $b$ arbitrary), and for ($i=j$; $a\neq b$)} \,\, .\label{Wijgluon}
\eea
\end{subequations}
Taking into account the trace over flavor space, which is the same as in the
polar phase, we obtain
\begin{enumerate}
\renewcommand{\labelenumi}{(\alph{enumi})}
\item $\m=\n=0$.
\be
\Pi_{ab}^{00}(0)=\d_{ab}\frac{g^2}{18\p^2}
\sum_{n=1}^{N_f}\m_n^2
\left\{
\begin{array}{cl}  3(v^{12}+v^{21})+3v^{22}  & \\
+6(w^{12}+w^{21})-3w^{22} &  \quad \mbox{for} \quad 
a=1,3,4,6,8 \,\, ,\\ \\
2v^{11}+(v^{12}+v^{21})+5v^{22} & \\
+8w^{11}-2(w^{12}+w^{21})+5w^{22}
&  \quad \mbox{for} \quad a=2,5,7 \,\, .\\
\end{array} \right.
\ee

\item $\m=0,\n=i$ and $\m=i,\n=0$.
\be
\Pi_{ab}^{0i}(0) = \Pi_{ab}^{i0}(0) = 0 \,\, .
\ee 

\item $\m=i,\n=j$. The general result (keeping all functions $v$, $w$)
is a complicated $24\times 24$ matrix and therefore not shown here.
However, as stated above, the polarization tensor 
is diagonal after inserting the values for the functions $v$ and $w$, 
and the result for the Debye and Meissner masses will be given in 
the next section. 

\end{enumerate}

\subsubsection{Mixed polarization tensor ($a\le 8, b=9$ and $a=9, b\le 8$)}

Here we have 
\begin{subequations} \label{VWCSLmixed}
\bea 
{\cal V}_{a\g,e_1e_2}^{\m\n,rs}
&=&\frac{eg}{2}\,\int\frac{d\Omega_{\bf k}}{(2\p)^3}
\Big\{ {\rm Tr}\left[\g^\m\,\g_0\,T_a\,{\cal P}_{\bf k}^r\,
\Lambda_{\bf k}^{-e_1}\,\g^\n\,\g_0\,Q\,{\cal P}_{\bf k}^s
\Lambda_{\bf k}^{-e_2}\right] + \left(e_{1,2}\to -e_{1,2},\; 
T_a\to T_a^T \right) \Big\}\,\, , \\
{\cal W}_{a\g,e_1e_2}^{\m\n,rs}&=&-\frac{eg}{2}\,
\int\frac{d\Omega_{\bf k}}{(2\p)^3}
\Big\{{\rm Tr}\left[\g^\m\,T_a\,{\bf J}\cdot(\uk-\gperp(\uk))
\,{\cal P}_{\bf k}^r\,\Lambda_{\bf k}^{e_1}\,\g_\n\,Q\,{\bf J}
\cdot(\uk-\gperp(\uk))\,{\cal P}_{\bf k}^s\,\Lambda_{\bf k}^{-e_2}
\right] \nonumber \\
&& \hspace{2.2cm} + \left(e_{1,2}\to -e_{1,2},\; T_a\to T_a^T \right)
\Big\}  \,\, ,
\eea
\end{subequations}
where $Q={\rm diag}(q_1,\ldots,q_{N_f})$.
All $(00)$, $(i0)$, and $(0i)$ components of these integrals vanish, 
\be \label{VWmixedvanish} 
{\cal V}^{00}_{a\g}={\cal V}^{0i}_{a\g}={\cal V}^{i0}_{a\g}
={\cal W}^{00}_{a\g}={\cal W}^{i0}_{a\g}={\cal W}^{0i}_{a\g}=0 \,\,.
\ee
(Here, we again omitted the indices $r$, $s$ and $e_1$, $e_2$.)
The $(ij)$  components of ${\cal V}$ and ${\cal W}$ are nonvanishing and, 
as for the gluonic CSL case,
we do not show them explicitly. However, the final result, i.e. the 
polarization tensor in the considered limit, vanishes
because of the following relations,
\begin{subequations} \label{VWijmixed} 
\bea
\sum_{rs}{\cal V}^{ij,rs}_{a\g,e_1,e_2}&=&0  \,\, , \label{Vijmixed} \\
{\cal W}_{a\g,e_1e_2}^{ij,rs}&\sim& 1-e_1e_2 \,\, . \label{Wijmixed}
\eea
\end{subequations}

\begin{enumerate}
\renewcommand{\labelenumi}{(\alph{enumi})}
\item $\m=\n=0$.
According to Eq.\ (\ref{VWmixedvanish}), we have
\be \label{CSLmixed00}
\Pi_{a\g}^{00}(0)=0\,\, .
\ee
\item $\m=0,\n=i$ and $\m=i,\n=0$.
\be
\Pi_{a\g}^{0i}(0) = \Pi_{a\g}^{i0}(0) = 0 \,\, .
\ee 
\item $\m=i,\n=j$. This polarization tensor  
has a complicated structure in terms of $v$, $w$ and is thus not given here.
However, as stated above, the final result is zero. 

\end{enumerate}

\subsubsection{Photon polarization tensor ($a=b=9$)}

For the photon polarization tensor in the CSL phase we need
\begin{subequations} \label{VWCSLphoton}
\bea 
{\cal V}_{\g\g,e_1e_2}^{\m\n,rs}
&=&\frac{e^2}{2}\,\int\frac{d\Omega_{\bf k}}{(2\p)^3}
\Big\{ {\rm Tr}\left[\g^\m\,\g_0\,Q\,{\cal P}_{\bf k}^r\,
\Lambda_{\bf k}^{-e_1}\,\g^\n\,\g_0\,Q\,{\cal P}_{\bf k}^s
\Lambda_{\bf k}^{-e_2}\right] + \left(e_{1,2}\to -e_{1,2}\right) 
\Big\}\,\, , \\
{\cal W}_{\g\g,e_1e_2}^{\m\n,rs}&=&
-\frac{e^2}{2}\,\int\frac{d\Omega_{\bf k}}{(2\p)^3}
\Big\{ {\rm Tr}\left[\g^\m\,Q\,{\bf J}\cdot(\uk-\gperp(\uk))\,
{\cal P}_{\bf k}^r\,\Lambda_{\bf k}^{e_1}\,\g_\n\,Q\,{\bf J}
\cdot(\uk-\gperp(\uk))\,{\cal P}_{\bf k}^s\,\Lambda_{\bf k}^{-e_2}\right] 
\nonumber \\
&& \hspace{2.2cm} + \left(e_{1,2}\to -e_{1,2}\right) 
\Big\} \,\, .
\eea
\end{subequations}
The results for the $(00)$ and $(ij)$ 
components are given in Tables \ref{tableCSL2} and \ref{tableCSL3}.
The $(0i)$ and $(i0)$ components vanish (for all $r$, $s$, $e_1$, $e_2$),
\be
{\cal V}_{\g\g}^{i0}={\cal V}_{\g\g}^{0i}=
{\cal W}_{\g\g}^{i0}={\cal W}_{\g\g}^{0i}=0 \,\, .
\ee
\begin{table}
\begin{tabular}{|c|c|c|}
\hline
${\cal V}_{\g\g,e_1e_2}^{00,11}$ & ${\cal V}_{\g\g,e_1e_2}^{00,12}=
{\cal V}_{\g\g,e_1e_2}^{00,21}$ & ${\cal V}_{\g\g,e_1e_2}^{00,22}$ \\
\hline
\hline
$\,\,(1+e_1e_2)/(2\p^2)\,\,$ & 0 & $(1+e_1e_2)/\p^2$ \\
\hline
\hline
${\cal W}_{\g\g,e_1e_2}^{00,11}$ & 
$\,\,{\cal W}_{\g\g,e_1e_2}^{00,12}={\cal W}_{\g\g,e_1e_2}^{00,21}\,\,$ & 
${\cal W}_{\g\g,e_1e_2}^{00,22}$ \\
\hline
\hline
$\,\,-2(1+e_1e_2)/\p^2\,\,$ & 0 & $\,\,-(1+e_1e_2)/\p^2\,\,$ \\
\hline
\end{tabular}
\caption{$(00)$ components of the tensors ${\cal V}$, ${\cal W}$ for the
photon polarization tensor in the CSL phase, defined in 
Eqs.\ (\ref{VWCSLphoton}).  }
\label{tableCSL2}
\end{table}
\begin{table}
\begin{tabular}{|c|c|c|}
\hline
${\cal V}_{\g\g,e_1e_2}^{ij,11}$ & ${\cal V}_{\g\g,e_1e_2}^{ij,12}=
{\cal V}_{\g\g,e_1e_2}^{ij,21}$ & ${\cal V}_{\g\g,e_1e_2}^{ij,22}$ \\
\hline
\hline
$\d^{ij}(11+7e_1e_2)/(54\p^2)$ & $8\,\d^{ij}(1-e_1e_2)/(27\p^2)$  & 
$\d^{ij}(19-e_1e_2)/(27\p^2)$ \\
\hline
\hline
${\cal W}_{\g\g,e_1e_2}^{ij,11}$ & 
${\cal W}_{\g\g,e_1e_2}^{ij,12}={\cal W}_{\g\g,e_1e_2}^{ij,21}$ & 
${\cal W}_{\g\g,e_1e_2}^{ij,22}$ \\
\hline
\hline
$\,\,2\,\d^{ij}(7+11e_1e_2)/(27\p^2)\,\,$ & 
$\,\,16\,\d^{ij}(1-e_1e_2)/(27\p^2)\,\,$ & 
$\,\,-\d^{ij}(1-19e_1e_2)/(27\p^2)\,\,$ \\
\hline
\end{tabular}
\caption{$(ij)$ components of the tensors ${\cal V}$, ${\cal W}$ for the
photon polarization tensor in the CSL phase, defined in 
Eqs.\ (\ref{VWCSLphoton}).  }
\label{tableCSL3}
\end{table}
One obtains the following photon polarization tensor.
\begin{enumerate}
\renewcommand{\labelenumi}{(\alph{enumi})}
\item $\m=\n=0$. 
\be \label{CSLphoton00}
\Pi_{\g\g}^{00}(0)=\sum_{n=1}^{N_f}q_n^2\m_n^2\,
\frac{e^2}{\p^2}(v^{11}+2v^{22}-4w^{11}-2w^{22}) \,\, .
\ee
\item $\m=0,\n=i$ and $\m=i,\n=0$.
\be
\Pi_{\g\g}^{0i}(0) = \Pi_{\g\g}^{i0}(0) = 0 \,\, .
\ee 
\item $\m=i,\n=j$.
\bea 
\Pi_{\g\g}^{ij}(0)&=&\d^{ij}\sum_{n=1}^{N_f}q_n^2\m_n^2\,
\frac{e^2}{27\p^2}\left[9v^{11}+2\bar{v}^{11}+16(\bar{v}^{12}
+\bar{v}^{21})+18v^{22}+20\bar{v}^{22} \right. \nonumber \\
&&\left. \hspace{2.8cm} +\,36w^{11}-8\bar{w}^{11}+32(\bar{w}^{12}+\bar{w}^{21})
+18w^{22}-20\bar{w}^{22}\right] \label{CSLphotonij} \,\, .
\eea
\end{enumerate}

\section{Mixing and screening} \label{results}

In this section we use the results of the previous section to calculate
the Debye and Meissner masses. We insert 
the numbers from Table \ref{tablevw} into the results for the
polarization tensors $\Pi_{ab}^{00}(0)$ and $\Pi_{ab}^{ij}(0)$ given
in Secs.\ \ref{2SCphase}, \ref{CFLphase}, \ref{polarphase}, \ref{CSLphase}
and use the definitions of the screening masses given in 
Sec.\ \ref{propagators} (cf.\ Eqs. (\ref{defmasses}) and comment below this
equation). 

We distinguish between the normal-conducting and the superconducting
phase. The screening properties of the normal-conducting phase are obtained
with the numbers given in Table \ref{tablevw} for temperatures larger
than the critical temperature for the superconducting phase transition,
$T\ge T_c$. For all $a,b\le 9$,  they lead to a 
vanishing Meissner mass, i.e., as 
expected, there is no Meissner effect in the normal-conducting state.
However, there is electric screening for temperatures larger than $T_c$.
Here, the Debye mass solely depends on the number of quark flavors
and their electric charge. We find (with $a\le 8$)
\be \label{normaldebye}
T\ge T_c: \qquad m_{D,aa}^2=3\,N_f\frac{g^2\m^2}{6\p^2} \quad, \qquad m_{D,a\g}^2=0
\quad,\qquad m_{D,\g\g}^2=18\sum_n q_n^2 \frac{e^2\m_n^2}{6\p^2} \,\, .
\ee
Consequently, the $9\times 9$ Debye mass matrix is already diagonal. 
Electric gluons and electric photons are screened. 

The masses in the superconducting phases are more interesting.
The results for all phases are collected in Table \ref{tabledebye}
(Debye masses) and Table \ref{tablemeissner} (Meissner masses). The physically
relevant, or ``rotated'', masses are obtained after a 
diagonalization of the
$9\times 9$ mass matrices. We see from Tables \ref{tabledebye} and
\ref{tablemeissner} that the special situation discussed in 
Sec.\ \ref{mixing} applies to all considered phases; namely, all 
off-diagonal gluon masses, $a,b\le 8$, as well as all mixed masses
for $a\le 7$ vanish. Furthermore, in all cases where the mass matrix
is not diagonal, we find $m_{8\g}^2=m_{88}m_{\g\g}$. Therefore, 
the rotated masses $\tilde{m}_{88}$ and $\tilde{m}_{\g\g}$ (which are the
eigenvalues of the mass matrix) are determined by Eqs.\ (\ref{detvanish}).
The electric and magnetic mixing angles $\theta_D\equiv\theta_2$ and 
$\theta_M\equiv\theta_1$ are 
obtained with the help of Eqs.\ (\ref{cossintheta}). Remember that the indices
1 and 2 originate from the spatially transverse and longitudinal
projectors defined in Eqs.\ (\ref{defABE}). They were associated with
the Meissner and Debye masses in Eqs.\ (\ref{defmasses}). We collect
the rotated masses and mixing angles for all phases in 
Table \ref{tablemixing}.

\begin{table}
\begin{tabular}{|c||cccccccc|cc|c|}
\hline
& \multicolumn{8}{c}{$m^2_{D,aa}$}\vline & 
\multicolumn{2}{c}{$m^2_{D,a\g}=m^2_{D,\g a}$}\vline
&  $m_{D,\g\g}^2$ \\
\hline\hline
$a$ & 1 & 2 & 3 & 4 & 5 & 6 & 7 & 8 & 1-7 & 8 & 9 \\ 
\hline\hline
2SC & \multicolumn{3}{c}{0}\vline & 
\multicolumn{4}{c}{$\frac{3}{2}g^2$}
\vline &  $3\,g^2$ & 0  & 0  &  $2\,e^2$ \\
\hline
CFL  & \multicolumn{8}{c}{$3\,\z\,g^2$}\vline  
 & 0 &  $-2\,\sqrt{3}\,\z\,eg$ &  $4\,\z\,e^2$ \\
\hline
polar & \multicolumn{3}{c}{0}\vline & 
\multicolumn{4}{c}{$\frac{3}{2}g^2$}\vline 
& $3\,g^2$ & 0 & 0 &  $18\,q^2 e^2$ \\
\hline   
CSL & $\,\, 3\,\b g^2\,\,$ & $\,\, 3\,\a g^2\,\,$ & $\,\, 3\,\b g^2\,\,$ 
& $\,\, 3\,\b g^2\,\,$
 & $\,\, 3\,\a g^2\,\,$ & $\,\, 3\,\b g^2\,\,$ & $\,\, 3\,\a g^2\,\,$ & 
$\,\, 3\,\b g^2\,\,$ & 0 &  0 
& $18\,q^2 e^2$ \\
\hline 
\end{tabular} 
\caption{Zero-temperature Debye masses. All masses are given in units of 
$N_f\m^2/(6\p^2)$, where $N_f=2$ in the 2SC phase, $N_f=3$ in the CFL
phase, and $N_f=1$ in the polar and CSL phases. We use
the abbreviations $\z\equiv (21-8\ln 2)/54$, 
$\a\equiv (3+4\ln 2)/27$, and
$\b\equiv (6-4\ln 2)/9$. 
}\label{tabledebye}
\end{table}

\begin{table}
\begin{tabular}{|c||cccccccc|cc|c|}
\hline
& \multicolumn{8}{c}{$m^2_{M,aa}$}\vline & 
\multicolumn{2}{c}{$m^2_{M,a\g}=m^2_{M,\g a}$}\vline
&  $m_{M,\g\g}^2$ \\
\hline\hline
$a$ & 1 & 2 & 3 & 4 & 5 & 6 & 7 & 8 & 1-7 & 8 & 9 \\ 
\hline\hline
2SC  & \multicolumn{3}{c}{0}\vline & 
\multicolumn{4}{c}{$\frac{1}{2}g^2$}\vline  
 & $\frac{1}{3}g^2$ & 0 &  $\frac{1}{3\sqrt{3}}eg $ &  
$ \frac{1}{9}e^2$ \\
\hline
CFL  & \multicolumn{8}{c}{$\z\,g^2$}\vline & 0  &  
$-\frac{2}{\sqrt{3}}\,\z\,eg$  &  $\frac{4}{3}\,\z\,e^2$ \\
\hline
polar & \multicolumn{3}{c}{0}\vline & 
\multicolumn{4}{c}{$\frac{1}{2}g^2$}\vline & 
 $\frac{1}{3}g^2$ & 0 &   $\frac{2}{\sqrt{3}}\,q\,eg$ &  
$4\,q^2 e^2$ \\
\hline   
CSL & $\,\,\b g^2\,\,$ & $\,\,\a g^2\,\,$ & $\,\,\b g^2\,\,$ & $\,\,\b g^2\,\,$ & $\,\,\a g^2\,\,$ & $\,\,\b g^2\,\,$ & $\,\,\a g^2\,\,$ & $\,\,\b g^2\,\,$ & 0 &  0 
& $6 \, q^2 e^2$ \\
\hline 
\end{tabular} 
\caption{Zero-temperature Meissner masses. All results are given in the same units as the 
Debye masses in Table \ref{tabledebye}. The abbreviations of Table
\ref{tabledebye} are used.}
\label{tablemeissner}
\end{table}

\begin{table}
\begin{tabular}{|c||c|c|c|c|c|c|}
\hline
 & $\tilde{m}_{D,88}^2$ & $\tilde{m}_{D,\g\g}^2$ & $\cos^2\theta_D$ & 
$\tilde{m}_{M,88}^2$ &$\tilde{m}_{M,\g\g}^2$ & $\cos^2\theta_M$ \\
\hline
\hline
2SC & $3\,g^2$ & $2\,e^2$ & 1 & $\frac{1}{3}g^2+\frac{1}{9}e^2$ & 0 
 & $3g^2/(3g^2+e^2)$ \\
\hline
CFL & $\quad (4\,e^2+3\,g^2)\,\z\quad $ & 0 & $\quad 3g^2/(3g^2+4e^2)\quad$ & 
$\quad \left(\frac{4}{3}e^2+g^2\right)\,\z\quad $ & 0 & $3g^2/(3g^2+4e^2)$ \\
\hline
polar & $3\,g^2$ & $\quad 18\,q^2e^2\quad$ & 1 & 
$\quad \frac{1}{3}g^2+4\,q^2e^2\quad$ & 0 
& $\quad g^2/(g^2+12\,q^2e^2)\quad$ \\
\hline
CSL & $3\,\b\,g^2$ & $18\,q^2e^2$ & 1 & $\b\,g^2$ & $\quad 6\,q^2e^2\quad$ 
& 1 \\
\hline
\end{tabular}
\caption{Zero-temperature rotated Debye and Meissner masses in units of 
$N_f\m^2/(6\p^2)$
and mixing angles for electric and magnetic gauge bosons. The constants 
$\z$, $\a$, and $\b$ are defined as in Tables \ref{tabledebye} and
\ref{tablemeissner}.}
\label{tablemixing}
\end{table}

Let us first discuss the spin-zero cases, 2SC and CFL.   
In the 2SC phase, the gluon masses are obtained from 
Eqs.\ (\ref{2SCgluon00}) and (\ref{2SCgluonij}). 
Due to a cancellation of the normal and anomalous
parts, represented by $v$ and $w$, the Debye and Meissner masses for the
gluons 1,2, and 3 vanish. Physically, this is easy to understand. Since
the condensate picks one color direction, all quarks of the third color, say
blue, remain unpaired. The first three gluons only interact with red and 
green quarks and thus acquire neither a Debye nor a Meissner mass.
We recover the results of Ref.\ \cite{meissner2}.
For the mixed and photon masses we inserted the electric charges 
for $u$ and $d$ quarks, i.e., in 
Eqs.\ (\ref{2SCmixed00}), (\ref{2SCmixedij}), (\ref{2SCphoton00}), and
(\ref{2SCphotonij}) we set $q_1=2/3$ and $q_2=-1/3$.
Here we find the remarkable result that the mixing angle for the Debye
masses is different from that for the Meissner masses, $\theta_D\neq\theta_M$.
The Meissner mass matrix is not diagonal. By a rotation with the angle 
$\theta_M$, given in Table \ref{tablemixing}, we diagonalize this matrix 
and find a vanishing 
mass for the new photon. Consequently, there is no electromagnetic Meissner
effect in this case. This fact is well-known \cite{emprops}. 
The Debye mass matrix, however, is diagonal. The off-diagonal elements 
$m_{D,8\g}^2$ vanish, since the contribution of the ungapped modes, 
corresponding to 
the blue quarks, $v^{22}$, cancels the one of the gapped modes, 
$v^{11}-w^{11}$, cf.\ Eq.\ (\ref{2SCmixed00}). Consequently,
the mixing angle is zero, $\theta_D=0$. Physically, this means that 
not only the color-electric eighth gluon but also the electric photon
is screened. Had we considered only the gapped quarks, i.e., 
$v^{22}=0$ in Eqs.\ (\ref{2SCgluon00}), (\ref{2SCmixed00}), 
and (\ref{2SCphoton00}), we would have found the same mixing angle as for
the Meissner masses and a vanishing Debye mass for the new photon. This
mixing angle is the same as predicted from simple group-theoretical 
arguments, Eq.\ (\ref{allangles}). The photon Debye mass in the 
superconducting 2SC phase differs from that of the normal phase, 
Eq.\ (\ref{normaldebye}), which, for $q_1=2/3$ and $q_2=-1/3$ is
$m_{D,\g\g}^2=5\,N_fe^2\m^2/(6\p^2)$. 

In the CFL phase, all eight gluon Debye and Meissner masses are equal. This 
reflects the symmetry of the condensate where there is no preferred
color direction. For the mixed and photon masses, 
we used Eq.\ (\ref{uds}), i.e., 
we inserted the electric charges for $u$, $d$, and $s$ quarks into 
the more general expressions given in Sec.\ \ref{CFLphase}.
The results in Tables \ref{tabledebye} and \ref{tablemeissner} show
that both Debye and Meissner mass matrices have nonzero off-diagonal
elements, namely $m_{8\g}^2=m_{\g 8}^2$. Diagonalization yields a zero
eigenvalue in both cases. This means that neither electric nor magnetic
(rotated) photons are screened. Or, in other words, there is a charge
with respect to which the Cooper pairs are neutral. Especially, there
is no electromagnetic Meissner effect in the CFL phase, either. 
Note that the CFL phase is the only one considered in this paper in which 
electric photons are not screened.
Unlike the 2SC phase, both electric and magnetic gauge fields are rotated
with the same mixing angle $\theta_D=\theta_M$. This angle is well-known 
\cite{emprops,litim}. Note that, due to the spectrum of
the matrix $\g_0{\cal M}_{\bf k}^\dag{\cal M}_{\bf k}\g_0=
({\bf J}\cdot{\bf I})^2$, there are two gapped branches. Unlike the
2SC phase, there is no ungapped quasiparticle excitation branch. This is the
reason why both angles $\theta_D,\theta_M$ coincide with the one 
predicted in Eq.\ (\ref{allangles}).

Let us now discuss the spin-one phases, i.e., the polar and CSL phases.
For the sake of simplicity, all results in Tables \ref{tabledebye}, 
\ref{tablemeissner}, and \ref{tablemixing} refer to a single quark system,
$N_f=1$, where the quarks carry the electric charge $q$. After discussing
this most simple case, we will comment on the situation where $N_f>1$
quark flavors separately form Cooper pairs. The results for the gluon 
masses show that, up to a factor $N_f$, there is no difference between
the polar phase and the 2SC phase regarding screening of color fields. This
was expected since also in the polar phase the blue quarks remain 
unpaired. Consequently, the gluons with adjoint color index $a=1,2,3$ 
are not screened. Note that, due to Eq.\ (\ref{2SCpolar}), the spatial
$z$-direction picked by the spin of the Cooper pairs has no effect on the
screening masses. As in the 2SC phase, electric gluons do not mix with 
the photon. There is electromagnetic Debye screening, which, in this case,
yields the same photon Debye mass as in the normal phase, 
cf.\ Eq.\ (\ref{normaldebye}). The Meissner mass matrix is diagonalized by 
an orthogonal transformation defined by the mixing angle which equals 
the one in Eq.\ (\ref{allangles}). 

In the CSL phase, we find a special pattern of the gluon Debye and Meissner
masses. In both cases, there is a difference between the gluons corresponding
to the symmetric Gell-Mann matrices with $a=1,3,4,6,8$ and the ones
corresponding to the antisymmetric matrices, $a=2,5,7$. The reason for this
is, of course, the residual symmetry group $SO(3)_{c+J}$ that describes 
joint rotations
in color and real space and which is generated by a combination of the 
generators of the spin group $SO(3)_J$ and the antisymmetric Gell-Mann
matrices, $T_2$, $T_5$, and $T_7$. The remarkable property of the CSL phase
is that both Debye and Meissner mass matrices are diagonal. 
In the case of the Debye masses, the mixed entries of the matrix, $m_{D,a\g}$,
are zero because of the vanishing traces, Eq.\ (\ref{VWmixedvanish}), 
indicating that pure symmetry reasons are responsible for this fact 
(remember that, in the 2SC phase, the reason for the same fact was a
cancellation of the terms originating from the gapped and ungapped
excitation branches). There is a nonzero photon Debye mass which is identical
to that of the polar and the normal phase which shows that electric photons 
are screened in the CSL phase. Moreover, and only in this phase, 
also magnetic photons are screened. This means that 
there is an electromagnetic Meissner effect. Consequently, there is no
charge, neither electric charge,  nor color charge, nor any combination
of these charges, with respect to which the Cooper pairs are neutral. 
This was also shown in Sec.\ \ref{symmix} where we argued that in the
CSL phase there is no nontrivial residual $\tilde{U}(1)_{em}$, 
cf.\ Table \ref{tablesymmetry}. This feature of the CSL phase
was already discussed in Ref.\ \cite{schmitt2}. 

Finally, let us discuss the more complicated situation of a many-flavor 
system, $N_f>1$, which is in a superconducting state with spin-one
Cooper pairs. In both polar and CSL phases, this extension of the system 
modifies the results in Tables \ref{tabledebye}, \ref{tablemeissner}, and
\ref{tablemixing}. We have to include several different electric quark charges,
$q_1,\ldots,q_{N_f}$, and chemical potentials, $\m_1,\ldots,\m_{N_f}$, 
in a way explained below Eq.\ (\ref{2SCpolar}) and shown in the 
explicit results of the CSL phase, Sec.\ \ref{CSLphase}. In the CSL phase,
these modifications will change the numerical values of all masses, but 
the qualitative conclusions, namely that there is no mixing and electric as
well as magnetic screening, remain unchanged. In the case of the polar 
phase, however, a many-flavor system might change the conclusions 
concerning the Meissner masses. While in the one-flavor case, diagonalization
of the Meissner mass matrix leads to a vanishing photon Meissner mass, this
is no longer true in the general case with arbitrary $N_f$. There is 
only a zero eigenvalue if the determinant of the matrix vanishes, i.e.,
if $m_{M,8\g}^2=m_{M,88}\,m_{M,\g\g}$. Generalizing the results 
from Table \ref{tablemeissner}, this condition can be written as
\be \label{meissnersurface} 
\sum_{m,n}q_n(q_n-q_m)\,\m_n^2\m_m^2 =0\,\, .
\ee
Consequently, in general and for fixed charges $q_n$, there is a hypersurface
in the $N_f$-dimensional space spanned by the quark chemical potentials
on which there is a vanishing eigenvalue of the Meissner mass matrix 
and thus no electromagnetic Meissner effect. All remaining points in 
this space correspond to a situation where the (new) photon Meissner 
mass is nonzero (although, of course,
there might be a mixing of the eighth gluon and the photon). 
Eq.\ (\ref{meissnersurface}) is trivially fulfilled when the electric 
charges of all quarks are equal. Then we have no electromagnetic 
Meissner effect in the polar phase which is plausible since, regarding
electromagnetism, this situation is similar to the one-flavor case.
For specific values of the electric charges we find very simple 
conditions for the chemical potentials. 
In a two flavor system with $q_1=2/3$, $q_2=-1/3$, 
Eq.\ (\ref{meissnersurface}) reads
\be
\m_1^2\m_2^2=0 
\ee
while, in a three-flavor system with $q_1=-1/3$, $q_2=-1/3$, and $q_3=2/3$, 
we have
\be
(\m_1^2+\m_2^2)\m_3^2=0 \,\, .
\ee
Consequently, these systems {\it always}, i.e., for all combinations
of the chemical potentials $\m_n$,  exhibit the electromagnetic Meissner effect
in the polar phase except when they reduce to the above discussed
simpler cases (same electric charge of all quarks or a one-flavor system).

\section{Summary and Conclusions} \label{summary}

We have presented a calculation of the gluon and photon polarization tensors 
$\Pi_{ab}^{\m\n}(P)$, $a,b=1,\ldots,8,\g$, in a color superconductor 
which yield, 
in the zero-energy, low-momentum limit, the Debye and Meissner masses 
for the gauge bosons. These masses indicate in which cases electric and
magnetic fields are screened. Here, magnetic screening is equivalent to 
the Meissner effect. We have derived the gauge field propagators 
in a general treatment starting from the QCD partition function. It has been
shown that, in general, longitudinal and transverse modes of the gauge fields
are mixed in a nontrivial way. An ``unmixing'' transformation leads to a 
modified longitudinal polarization tensor. However, in the static limit,
this mixing is absent and longitudinal and transverse modes are well 
separated. 

We have shown that a nondiagonal polarization tensor
causes a mixing of the gauge fields. 
In general, the tensor has to be diagonalized via a transformation 
generated by an orthogonal operator ${\cal O}(P)$. Consequently,
also the gluon and photon fields, $A_\m^a$, are transformed. 
The operator ${\cal O}(P)$ defines new gauge fields, $\tilde{A}_\m^a$,
which are linear combinations of the original ones. This is in complete
analogy to the standard model of electroweak interactions, where, 
due to spontaneous breaking
of the electroweak symmetry $SU(2)\times U(1)$, the new fields
are the gauge bosons of the weak and electromagnetic interaction.
In this case, the three fields corresponding to the weak interaction
attain a mass via the Higgs mechanism. Similarly, also in the 
case of the color-superconducting phase transition, some of the 
originally nine gauge fields can become massive (besides a possible 
mixing). In the framework of the formalism presented here, the masses of
the new fields are the eigenvalues of the polarization tensor 
$\Pi^i_{ab}(0)$, where the index $i=1,2$ corresponds to the transverse and 
longitudinal modes (Meissner and Debye masses, respectively). 
Thus, the number of vanishing eigenvalues of $\Pi^i_{ab}(0)$ indicate the 
dimension of the unbroken subgroup of the color-electromagnetic 
product group $SU(3)_c\times U(1)_{em}$. 

We have presented a general formalism to calculate the polarization 
tensor for different color-superconducting phases. We have explicitly 
computed its zero-energy, low-momentum limit 
for the 2SC, CFL, polar, and CSL phases. Parts of the
results were already known in the literature, namely the gluon
Debye and Meissner masses for the spin-0 phases 2SC and CFL 
\cite{meissner2,meissner3,son}. Our result for the photon Debye mass
in the 2SC phase differs from that of Ref.\ \cite{litim} since our 
calculation shows that the photon-gluon mass matrix is already diagonal 
and thus 
electric gluons do not mix with the photon (while in Ref.\ \cite{litim}
the same diagonal matrix was rotated and a ``mixed mass'' was obtained).
The masses for the spin-1 phases have never been computed before. 
For the polar phase, we have shown that there is mixing between the 
magnetic gauge bosons but, as in the 2SC phase, no mixing of the electric 
gauge bosons. In a system of one quark flavor, this mixing leads to a
vanishing Meissner mass. However, for more than one quark flavor, we have 
shown that, if the electric charges of the quark flavors are not identical, 
there is an electromagnetic Meissner effect in the polar phase, contrary 
to both considered spin-0 phases. 
For the CSL phase, we have
found the remarkable result that, for any number of flavors, neither
electric nor magnetic gauge fields are mixed. Since there is no 
vanishing eigenvalue of $\Pi_{ab}^i(0)$, all eight gluons and the 
photon (electric as well as magnetic modes) become massive and 
there is an electromagnetic Meissner effect. 
This might affect the 
electromagnetic properties of a neutron star that has a core 
of color-superconducting quark matter in the CSL phase. We have already 
discussed these possible astrophysical implications  in 
Ref.\ \cite{schmitt2}.
There we also argued that, in spite of a suppression of the gap of 
three orders of magnitude compared to the spin-0 gaps \cite{schaefer,schmitt},
spin-1 gaps might be preferred in a charge-neutral system. The reason for 
that is that a mismatch of the Fermi surfaces of different quark flavors
has no effect on the spin-1 phases, where quarks of the same flavor
form Cooper pairs.

\section*{Acknowledgments}

The authors thank I.\ Shovkovy for interesting discussions. 
Q.W.\ acknowledges support from the Alexander von Humboldt-Foundation.

\appendix

\section{Integrals over quark momentum $k$} \label{AppA}

In this appendix, we prove the results shown in Table \ref{tablevw}, i.e.,
we calculate the integrals over quark momentum defined in 
Eqs.\ (\ref{kintegrals}). 
In this calculation we will neglect the antiparticle gap, 
$\phi^-\simeq 0$. 

\subsection{Contributions from normal propagators at $T=0$}

We start from Eqs.\ (\ref{defv}) which,
for $p_0=0$ reduce to (cf.\ Eq.\ (\ref{spurious}))
\be \label{v}
v_{e_1e_2}^{rs}=-\frac{\e_{1,r}\e_{2,s}-\x_1\x_2}{2\e_{1,r}\e_{2,s}
(\e_{1,r}+\e_{2,s})}
(1-N_{1,r}-N_{2,s})+\frac{\e_{1,r}\e_{2,s}+\x_1\x_2}{2\e_{1,r}\e_{2,s}
(\e_{1,r}-\e_{2,s})}
(N_{1,r}-N_{2,s}) \,\, .
\ee
Here, we inserted the definition of $n_{i,r}$ given in 
Eq.\ (\ref{abbreviations}) and defined
\be
\x_i\equiv e_ik_i-\m \,\, .
\ee
In the limit $p\to 0$, or, equivalently, $k_1\to k_2$, we obtain for 
$r=1,2$ 
\be
v^{rr}\simeq -\frac{1}{\m^2}\int_0^\infty dk\,k^2\left[
\frac{\l_r\phi^2}{4\e_r^3}(1-2N_r)-\frac{\e_r^2+\x^2}{2\e_r^2}
\frac{dN_r}{d\e_r}   \right] \,\, ,
\ee
where we used $(N_{1,r}-N_{2,r})/(\e_{1,r}-\e_{2,r})\simeq dN_r/d\e_r$
and $\xi\equiv k-\mu$.
All quantities now depend on $k_1=k$ and correspond to positive energies, 
$e_1=e_2=1$, since, due to $\phi^-\simeq 0$, the terms corresponding 
to negative energies vanish. Therefore, we omit the index $i=1,2$. 
Note that, up to subleading order,  the gap function does not depend 
on the index $r$, $\phi_1\simeq\phi_2\equiv\phi$ \cite{schmitt}. 
We have to distinguish between the cases where $\l_r\neq 0$ and 
where $\l_r=0$. When $\l_r\neq 0$, we obtain for zero temperature $T=0$,
where $N_r=0$, 
\be
v^{rr}=-\frac{1}{\m^2}\int_0^\infty dk\, k^2\frac{\l_r\phi^2}{4\e_r^3}
\simeq -\frac{1}{2}\int_0^\m d\x \l_r\phi^2(\x^2+\l_r\phi^2)^{-3/2}
\simeq -\frac{1}{2} \,\, .
\ee
Here, we restricted the $k$-integration to the region $0\le k\le 2\m$ since
the gap function is strongly peaked around the Fermi surface, $k=\m$.
In this region, we assume that the gap function does not depend on the
momentum $k$. 
Since the eigenvalue $\l_r$ cancels out, this result does not depend on 
$r$. Therefore it gives rise to the value of $v^{11}$ for all four
considered phases and for $v^{22}$ in the cases of the CFL and
CSL phases.
When $\l_r=0$, we obtain (for arbitrary temperature $T$)
\be
v^{rr}=\frac{1}{\m^2}\int_0^\infty dk \, k^2\frac{dN_F}{dk} \,\, ,
\ee 
where $N_F\equiv 1/[\exp(\x/T)+1]$ is the Fermi distribution. Using
$\m\gg T$, and the substitution $\z=\x/2T$, this can be transformed to
\be
v^{rr}\simeq -\int_0^\infty d\z \frac{1}{\cosh^2\z} = -1 \,\, .
\ee
Let us now compute $v^{rs}$ for $r\neq s$. Using Eq.\ (\ref{v}),
we obtain for two nonvanishing eigenvalues $\l_1,\l_2$ and at zero 
temperature
\be \label{hatv12}
v^{12}=v^{21}=-\frac{1}{\m^2} \int_0^\infty dk\, k^2
\frac{\e_1\e_2-\x^2}{2\e_1\e_2(\e_1+\e_2)}\simeq 
-\int_0^\m d\x \frac{1}{(\l_2-\l_1)\phi^2}\left(\frac{\e_2^2+\x^2}{\e_2}
-\frac{\e_1^2+\x^2}{\e_1}\right)\simeq -\frac{1}{2} \,\, .
\ee
Again, we neglected the integral over the region $k>2\m$.
In the last integral, the leading order contributions cancel. The 
subleading terms give rise to $(\l_2-\l_1)\phi^2/2$. Therefore, 
the eigenvalues cancel and the result does not depend on $\l_1,\l_2\neq 0$.
Eq.\ (\ref{hatv12}) holds for the CFL and CSL phases.

When one of the excitation branches is ungapped (2SC and polar phase), 
for instance, $\l_2=0$, we have
\be
v^{12}=v^{21}=-\frac{1}{\m^2}\int_0^\infty dk\,k^2 
\left[ \frac{n_1}{\e_1+\xi}
(1-N_1-N_F)-\frac{1-n_1}{\e_1-\x}(N_1-N_F)\right] \,\, .
\ee
For $T=0$, this becomes with the standard approximations
\bea
v^{12}&\simeq&-\frac{1}{\m^2}\int_{-\m}^\m d\x \frac{(\x+\m)^2}{2\e_1}
\left[\frac{\e_1-\x}{\e_1+\x}\Theta(\x)+\frac{\e_1+\x}{\e_1-\x}\Theta(-\x)
\right] \nonumber \\ 
&=&-\int_0^\m \frac{d\x}{\e_1}\left(1+\frac{\x^2}{\m^2}\right)
\frac{\e_1-\x}{\e_1+\x}\simeq -\frac{1}{2} \,\, ,
\eea
where the last integral has already been computed in Ref.\ \cite{meissner2}. 

Next we compute $\bar{v}^{rs}$ for the various cases. Setting the
thermal distribution function for antiparticles to zero, 
we obtain with the definition in Eq.\ (\ref{kintegrals})
\be  \label{tildevrs}
\bar{v}^{rs}=-\frac{1}{\m^2}\int_0^\infty dk\, k^2\left[
\frac{(1-n_r)(1-N_r)}{\e_r^++\e_s^-}-\frac{N_rn_r}{\e_r^+-\e_s^-}
+\frac{(1-n_s)(1-N_s)}{\e_r^-+\e_s^+}-\frac{N_sn_s}{\e_s^+-\e_r^-}\right]
\ee
with $\e_r^\pm\equiv\e_{k,r}^\pm$. Again, we first consider the situation where
$r=s$. In this case, for $T=0$ and $\l_r\neq 0$, (defining $\e\equiv\e^+$)
\bea \label{tildevfinal}
\bar{v}^{rr}&=&-\frac{2}{\m^2}\int_0^\infty dk\, k^2\left[\frac{1-n_r}{\e_r
+k+\m}-\frac{1}{2k}\right] \nonumber \\
&&=\frac{1}{\m^2}\int_0^\infty dk\, k \, \frac{\m(\e_r+\m-k)+\l_r\phi^2}
{\e_r(\e_r+k+\m)}\simeq \frac{1}{2} \,\, .
\eea
Here, $-1/(2k)$ is a vacuum subtraction. The last integral has been computed
in Ref.\ \cite{meissner2}. Note that the result does not depend on the value
of $\l_r\neq 0$. Thus, it is valid for $\bar{v}^{11}$ in the 2SC and 
polar phases as well as for both $\bar{v}^{11}$ and $\bar{v}^{22}$ in
the CFL and CSL phases.

For $\l_r= 0$, we obtain from Eq.\ (\ref{tildevrs}) at zero temperature
and with the vacuum subtraction $1/(2k)$
\be \label{tildevfinal2}
\bar{v}^{rr}=-\frac{2}{\m^2}\int_0^\infty dk\, k^2\left[\frac{\Theta(k-\m)}
{2k}-\frac{1}{2k}\right]=\frac{1}{2} \,\, .
\ee
This result holds for $\bar{v}^{22}$ in the 2SC and polar phases.

Next, we discuss the case $r\neq s$.  From Eqs.\ (\ref{tildevrs}) and 
(\ref{tildevfinal}) it is obvious, since the result for $\bar{v}^{rr}$ did
not depend on $\l_r$, that we get the same result for $\bar{v}^{rs}$ 
with two nonvanishing eigenvalues $\l_1,\l_2 \neq 0$. Thus, in this
case,
\be \label{tildev12}
\bar{v}^{12}=\bar{v}^{21}=\frac{1}{2} \,\, .
\ee
For $\l_2=0$, we obtain at zero temperature
\be
\bar{v}^{12}=\bar{v}^{21}=-\frac{1}{\m^2}\int_0^\infty dk\, k^2\left[
\frac{1-n_1}{\e_1+k+\m}+\frac{\Theta(k-\m)}{2k}-\frac{3}{2k}\right]
\simeq \frac{1}{2} \,\, ,
\ee
where identical integrals as in Eqs.\ (\ref{tildevfinal}) and
(\ref{tildevfinal2}) were performed. Consequently, also for the 2SC and 
polar phases, Eq.\ (\ref{tildev12}) holds.

\subsection{Contributions from anomalous propagators at $T=0$}

We start from Eq.\ (\ref{defw}) which, for $p_0=0$, is 
\be \label{wrs}
w_{e_1e_2}^{rs}=\frac{\phi^{e_1}\phi^{e_2}}{2\e_{1,r}\e_{2,s}(\e_{1,r}
+\e_{2,s})}(1-N_{1,r}-N_{2,s})+\frac{\phi^{e_1}\phi^{e_2}}
{2\e_{1,r}\e_{2,s}(\e_{1,r}-\e_{2,s})}(N_{1,r}-N_{2,s}) \,\, .
\ee
Obviously, since $\phi^-\simeq 0$, we have for all $r$, $s$ and all phases
$\bar{w}^{rs}=0$ (for the definition of $\bar{w}^{rs}$ 
cf.\ Eq.\ (\ref{kintegrals})).
First, we calculate $w^{rs}$ for $r=s$. Neglecting the antiparticle 
gap and taking the limit $k_1\to k_2$, we obtain, 
using the same notation as above, 
$\e_r\equiv \e^+_{k,r}$, 
\be 
w^{rr}=\frac{1}{\m^2}\int_0^\infty dk \, k^2\left[\frac{\phi^2}{4\e_r^3}
(1-2N_r)+\frac{\phi^2}{2\e_r^2}\frac{dN_r}{d\e_r}\right] \,\, .
\ee
For $\l_r\neq 0$ and at zero temperature, where $N_r=0$, this expression 
reduces to
\be
w^{rr}=\frac{1}{\m^2}\int_0^\infty dk \, k^2 \frac{\phi^2}{4\e_r^3}
\simeq\frac{1}{2}\int_0^\m\d\x\,\phi^2(\x^2+\l_r\phi^2)^{-3/2}\simeq
\frac{1}{2\l_r} \,\, .
\ee
Unlike in $v^{rr}$, the eigenvalue $\l_r$ does not cancel out and we
obtain different results for $\l_r=4$ and $\l_r=1$. In the cases of 
the 2SC and polar phases we obtain $w^{11}=1/2$ whereas for
the CFL and CSL phases, $w^{11}=1/8$ and $w^{22}=1/2$.
In the former two cases, the quantity $w^{22}$ does not occur
in our calculation. 
Thus we can turn to the case where $r\neq s$. Here, we conclude from 
Eq.\ (\ref{wrs}) 
\be
w^{12}=w^{21}=\frac{1}{\m^2}\int_0^\infty dk \, k^2 
\left[\frac{\phi^2}{2\e_1\e_2(\e_1+\e_2)}(1-N_1-N_2)+ 
\frac{\phi^2}{2\e_1\e_2(\e_1-\e_2)}(N_1-N_2)\right] \,\, .
\ee
For two nonvanishing eigenvalues $\l_1$, $\l_2$, this reads at $T=0$
\bea
w^{12}&=&\frac{1}{\m^2}\int_0^\infty dk \, k^2 \frac{\phi^2}
{2\e_1\e_2(\e_1+\e_2)}\simeq\frac{1}{\l_1-\l_2}
\int_0^\m d\x \, \left(\frac{1}{\e_2}-\frac{1}{\e_1}\right) \nonumber \\
&\simeq&\frac{1}{2}\frac{1}{\l_1-\l_2}\ln\frac{\l_1}{\l_2} \,\, .
\eea
For $\l_1=4$ and $\l_2=1$ (CFL and CSL phases) we obtain 
\be
w^{12}=w^{21}=\frac{1}{3}\ln 2 \,\, .
\ee
The corresponding expression for the case where the second eigenvalue 
vanishes, $\l_2=0$, does not occur in our calculation.

\subsection{Integrals in the normal phase, $T\ge T_c$}

For temperatures larger than the transition temperature $T_c$ but still
much smaller than the chemical potential, $T\ll \m$, all integrals defined
in Eqs.\ (\ref{kintegrals}) are easily computed. Since there is no 
gap in this case, $\phi = 0$, all contributions from the anomalous 
propagators vanish trivially, $w^{rs}=\bar{w}^{rs}=0$. From 
Eq.\ (\ref{v}), which hold for all temperatures, we find with
$\phi = 0$
\be
v^{11}=v^{22}=v^{12}=\frac{1}{\m^2}\int_0^\infty dk\, k^2
\frac{dN_F}{d\x}\simeq -1 
\ee
and (with the vacuum subtraction $1/k$)
\be
\bar{v}^{11}=\bar{v}^{22}=\bar{v}^{12}=-\frac{1}{\m^2}
\int_0^\infty dk\, k^2 \left[\frac{1-N_F}{k}-\frac{1}{k}\right]
\simeq \frac{1}{2} \,\, .
\ee

\end{document}